\documentclass[11pt,a4paper]{article}
\usepackage{jheppub,amsmath,  amssymb,slashed,url,bm,textgreek,slashed,upgreek,relsize}
\usepackage{graphicx}
\usepackage{epstopdf}
\UseRawInputEncoding
\def\t{{ \sf t}} 
\def\max{{\mathrm{max}}}

\def\WW{{\mathcal W}}

\def\Q{{\sf Q}}

\def\DD{{\mathcal D}}

\def\gf{{\mathrm{gf}}}

\def\Chi{\chi}

\def\be{\begin{equation}}
\def\ee{\end{equation}}

\def\hat{\widehat}

\def\h{\widehat}

\def\D{{\sf{Diff}}}
\def\S{{\mathcal S}}

\def\bigdelta{{\mathlarger{\mathlarger\updelta}}}
\def\largedelta{{\mathlarger{\updelta}}}

\def\O{{\mathcal O}}

\def\gh{{\mathrm{gh}}}

\def\d{{\mathrm d}}

\def\Ca{{\sf{Conf}}}

\def\b{\overline}
\def\diff{{\mathrm{diff}}}
\def\R{{\mathbb R}}
\def\C{{\mathbb C}}

\def\Diff{{\sf{Diff}}}
\def\[{\bigl [}

\def\]{\bigr ]}

\def\Z{{\mathbb Z}}

\def\t{\widetilde }
\def\h{\widehat}

\def\K{{\sf K}}

\def\I{{\mathcal I}}

\def\G{{\mathcal G}}

\def\x{{\vec x}}

\def\M{{\sf{Met}}}
\def\matt{{\mathrm{matt}}}
\def\Met{{\sf{Met}}}
\def\Conf{{\sf{Conf}}}

\def\HH{{ H}}
\def\W{{\sf{Weyl}}}
\def\P{{\mathcal P}}

\def\phys{{\mathrm{phys}}}

\def\H{{\mathcal H}}

\def\i{{\mathrm i}}

\def\bar{\overline}
\def\DD{{\mathfrak D}}

\def\G{{\sf G}}
\def\past{{\mathrm{past}}}
\def\fut{{\mathrm{fut}}}
\def\AdS{{\mathrm{AdS}}}

\def\vP{\varPhi}

\title{A Note On The Canonical Formalism For Gravity}

\def\la{\langle}
\def\ra{\rangle}
\def\S{{\mathcal S}}

 \author{Edward Witten}
\affiliation{School of Natural Sciences, Institute for Advanced Study,\\ 1 Einstein Drive, Princeton, NJ 08540 USA}
\abstract{We describe a simple gauge-fixing that leads to a construction of a quantum Hilbert space for quantum gravity in an asymptotically Anti de Sitter
spacetime, valid to all orders of perturbation theory.  The construction is motivated by a relationship of the phase space of gravity in asymptotically Anti de Sitter
spacetime to a cotangent bundle.   We describe what is known about this relationship and some extensions that  might plausibly be true. 
A key fact is  that, under certain conditions,  the Einstein Hamiltonian constraint equation
can be viewed as a way to gauge fix the group of conformal rescalings of the metric of a Cauchy hypersurface.
 An analog of the procedure that
we follow for Anti de Sitter gravity leads to standard results for a Klein-Gordon particle.
}

\begin{document}\maketitle
\tableofcontents
\section{Introduction}\label{intro} 
In this article, we will re-examine the canonical formalism for quantum gravity \cite{DeWitt}, focusing on the case of an asymptotically Anti de Sitter (AAdS) spacetime $X$.
One advantage of the AAdS case is that, because of holographic duality, it is possible to explain in a straightforward way what problem the canonical formalism
is supposed to solve, thereby circumventing questions like what observables to consider and what is a good notion of ``time.''  
In holographic duality, there is a straightforward notion of boundary time, and there is no difficulty in defining local boundary observables.

\begin{figure}
 \begin{center}
   \includegraphics[width=4in]{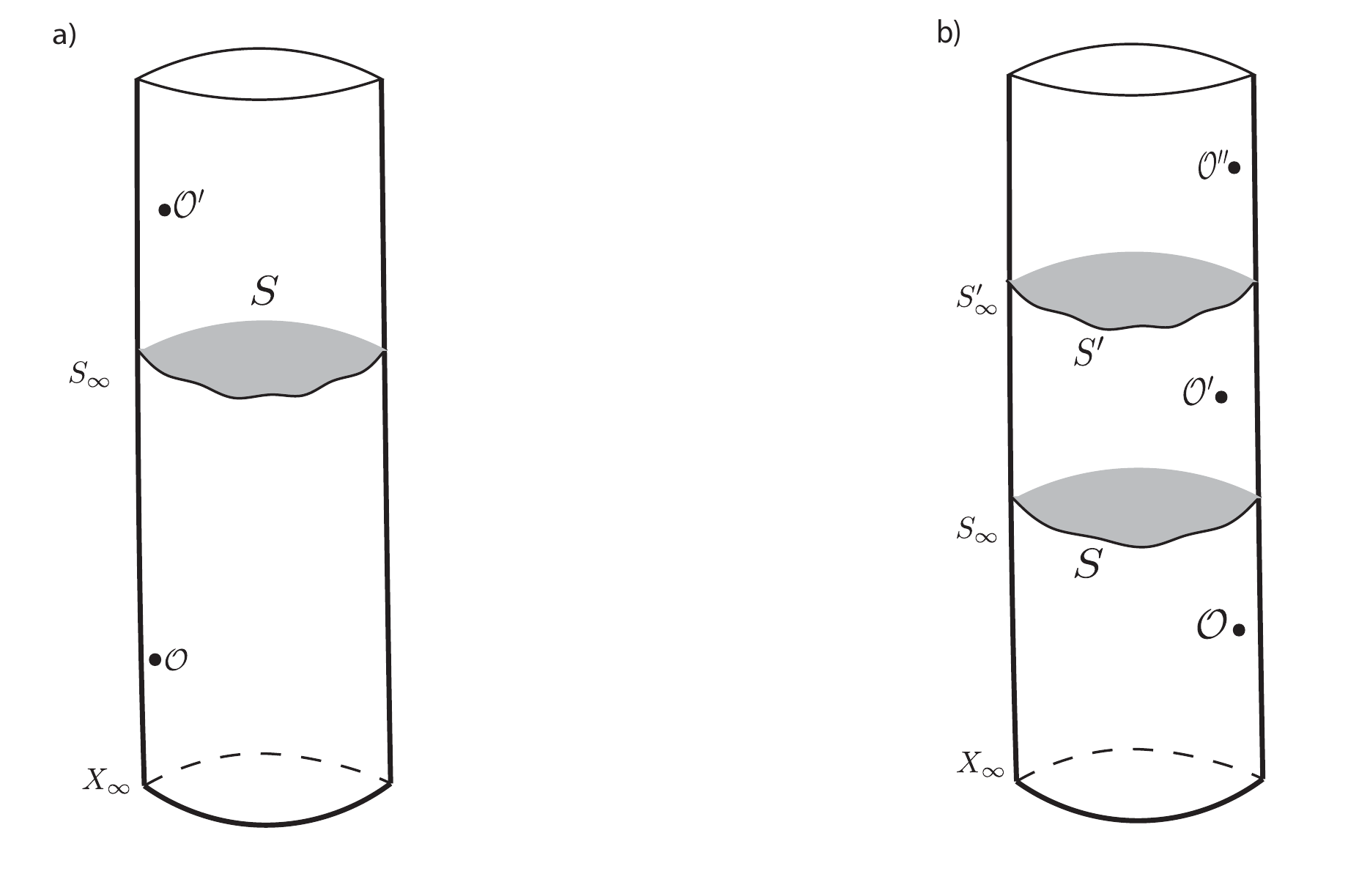}
 \end{center}
\caption{\small (a) The boundary $X_\infty$ of Anti de Sitter space, drawn as a cylinder, with 
 boundary  insertions $\O$ and $\O'$ to the past and future of  a Cauchy hypersurface $S_\infty\subset X_\infty$.   In this setup, one can compute a matrix element
$\la \Psi|\O'(t',\vec x')\O(t,\vec x)|\chi\ra$ as a sum over physical states defined on the hypersurface $S_\infty$.   In the canonical formalism for gravity,
one aims to find a similar formula in terms of a sum over states on a bulk Cauchy hypersurface $S$ with boundary $S_\infty$.   (b)   The ``cutting'' procedure
of (a) can be iterated, with successive cuts on successive hypersurfaces.
 \label{BOX}}
\end{figure}

It is natural in holographic duality to study  the matrix elements of a product of local boundary operators $\O_\alpha$ between given initial and final states.
A typical example is  $\la \Psi|\O'(t',\vec x')\O(t,\vec x)|\chi\ra$, with boundary insertions of local operators $\O'$, $\O$ at points labeled by time $t$ and spatial coordinates 
 $\vec x$, and with states $\chi$, $\Psi$ that are defined
by initial and final  conditions.  For simplicity, in this article we restrict to $t'>t$, to avoid having to discuss ``timefolds.''  
 In the canonical formalism of the boundary theory, one constructs for any Cauchy hypersurface $S_\infty$ in the boundary of $X$
 a Hilbert space $\HH$ of quantum states with the property that if  some set $\I$ labels a basis $|i\ra$ of $\HH$, 
 then an amplitude can be factored by inserting a sum over these  states  (fig. \ref{BOX}(a)):
\be\label{formula}\la \Psi|\O'(t',\vec x')\O(t,\vec x)|\chi\ra=\sum_{i\in \I}\la\Psi|\O'(t',\vec x')|i\ra\la i|\O(t,\vec x)|\chi\ra. \ee
This factorization is most naturally described in path integrals  if $\O$ is to the past of $S_\infty$ and $\O'$ is to the future. 
Such factorization can be iterated; for example,  given two boundary Cauchy hypersurfaces $S_\infty$ and $S'_\infty$ with $S'_\infty$ to the future of $S_\infty$, and
insertions on the boundary arranged in time in a suitable way, one has (fig. \ref{BOX}(b))
\be\label{formula2}\la \Psi|\O''(t'',\vec x'')\O'(t',\vec x')\O(t,\vec x)|\chi\ra=\sum_{j'\in\I',\, i\in \I}\la\Psi|\O''(t'',\vec x'') |j'\ra\la j'|\O'(t',\vec x')|i\ra\la i|\O(t,\vec x)|\chi\ra \ee
where the states $|i\ra$ are defined on $S_\infty$ and the states $|j'\ra$ are defined on $S'_\infty$.    In writing the formula this way, we allow for the use of
different bases $\I$ on $\S_\infty$ and $\I'$ on $S'_\infty$.
The initial and final states $\chi$ and $\Psi$ in these formulas  are themselves  Hilbert space states defined in the far past and the far future.

The main result
 of the present article is a conceptually simple way to reproduce such factorization laws from a bulk point of view, to all orders of perturbation theory.  The main idea is to exploit
 a relationship between Weyl invariance of a $(D-1)$-geometry and the Hamiltonian constraint equation of General Relativity.    
  
 Conformal approaches
 to quantization of gravity have a very long history \cite{York}, and the conformal approach to the constraint equations, which gives particularly simple
 results in the case of an AAdS spacetime, has been much developed \cite{OY,Ch,CI,I,crusc,BI}.   As we will see, the  conformal approach is particularly powerful when it
 can be combined with existence and
 uniqueness results for maximal volume hypersurfaces, as was done for three-dimensional pure gravity in \cite {Monc,KS,BoS,SK}.

 In section \ref{gravh}, we explain a bare minimum of this classical picture to motivate the approach that
we will take to the canonical formalism of gravity.    Then we go on to describe, by a simple gauge-fixing, a construction of a Hilbert space  that is valid to all orders of 
perturbation theory.      In section \ref{class}, we explain the underlying classical picture more thoroughly.   

In early investigations of the canonical formalism for gravity \cite{DeWitt}, it was observed that the Hamiltonian constraint of General Relativity is
a family of 
 second order differential operators,  somewhat analogous to a Klein-Gordon operator.   This suggested that the inner product for gravity might be defined
by analogy with a   Klein-Gordon bilinear 
pairing $(f,g)=\int_S \d \Sigma^\mu \bar f \overset{\leftrightarrow} {\partial_\mu}g$, which does not depend on the choice of the hypersurface $S$ on which it
is evaluated.   The analogy has always seemed problematical, because the Klein-Gordon pairing is not positive-definite, and also because the Hamiltonian
constraint is a whole infinite family of second order operators, not just one.  We will see that the procedure we follow
for gravity, though it leads to a positive inner product, is closely analogous to a procedure which for a Klein-Gordon particle leads to the Klein-Gordon pairing.

Our analysis is restricted to perturbation theory  for  technical reasons, and  it may  be that this is inherent in assuming that $\HH$ can be constructed as a space
of functions of fields -- the metric tensor and possibly other fields -- on a  spacetime manifold.  
However, the result we get for the Klein-Gordon particle is exact, even though the derivation appears to be valid only in perturbation theory.   
In addition, the classical picture that motivates the present work is valid far beyond what is needed for perturbation theory.
These facts suggest that in some admittedly unclear sense, the description of canonical quantization given in the present article
might extend beyond perturbation theory.  This possibility  has motivated the writing of section \ref{class} of the present article.   Much of that section is an explanation
of the conformal approach to the classical constraint equations, largely following the useful review article \cite{crusc}.

An early version of this work, but without the gauge-fixing construction of section \ref{gaugefixing}, was presented in a lecture at the Princeton Center for Theoretical
Science \cite{WittenLecture}.

As already noted, the main idea in this article is to exploit a relationship between the Hamiltonian constraint equation of gravity and the group of Weyl rescalings
of a Cauchy hypersurface.   Another and arguably much deeper relationship between the Hamiltonian constraint and Weyl invariance has been
developed in recent years.   The $T\bar T$ deformation is a deformation of a two-dimensional quantum field theory that is irrelevant in the renormalization group
sense and for which no ultraviolet completion is understood 
but that nonetheless leads to unexpected exact results \cite{Zam,ZamTwo,Cardy}.    The Wheeler-DeWitt equation (or the Hamiltonian constraint
equation) of three-dimensional gravity without matter fields can be interpreted in terms of the $T\bar T$ deformation of two-dimensional conformal field theory
\cite{Verl}.   This striking insight has been refined and generalized to higher dimensions, where the $T\bar T$ deformation becomes a $T^2$ deformation \cite{KLM, Taylor,HKST,Shyam,ALS}.  More recently and 
remarkably, it  has been extended to include matter fields \cite{AKW}.   In the present article, the $T^2$ deformation plays no role in the input, but in a sense
we run into the $T^2$ deformation in the output, since the formula that we get for the Hilbert space inner product of a theory with gravity involves a sort
of $T^2$-deformed ghost determinant.    

The approach in the present article is limited to perturbation theory because we will use a gauge-fixing condition that is only valid  perturbatively.   The approach via
the $T^2$ deformation is, at the present time, limited to perturbation theory because a nonperturbative completion of the $T^2$ deformation is not known.   
As already noted, a limitation to perturbation theory may well be unavoidable in any description based on fields in spacetime.

\section{Path Integrals and Physical States}\label{gravh}

\subsection{The Phase Space As A Cotangent Bundle}\label{orient}

 \begin{figure}
 \begin{center}
   \includegraphics[width=2in]{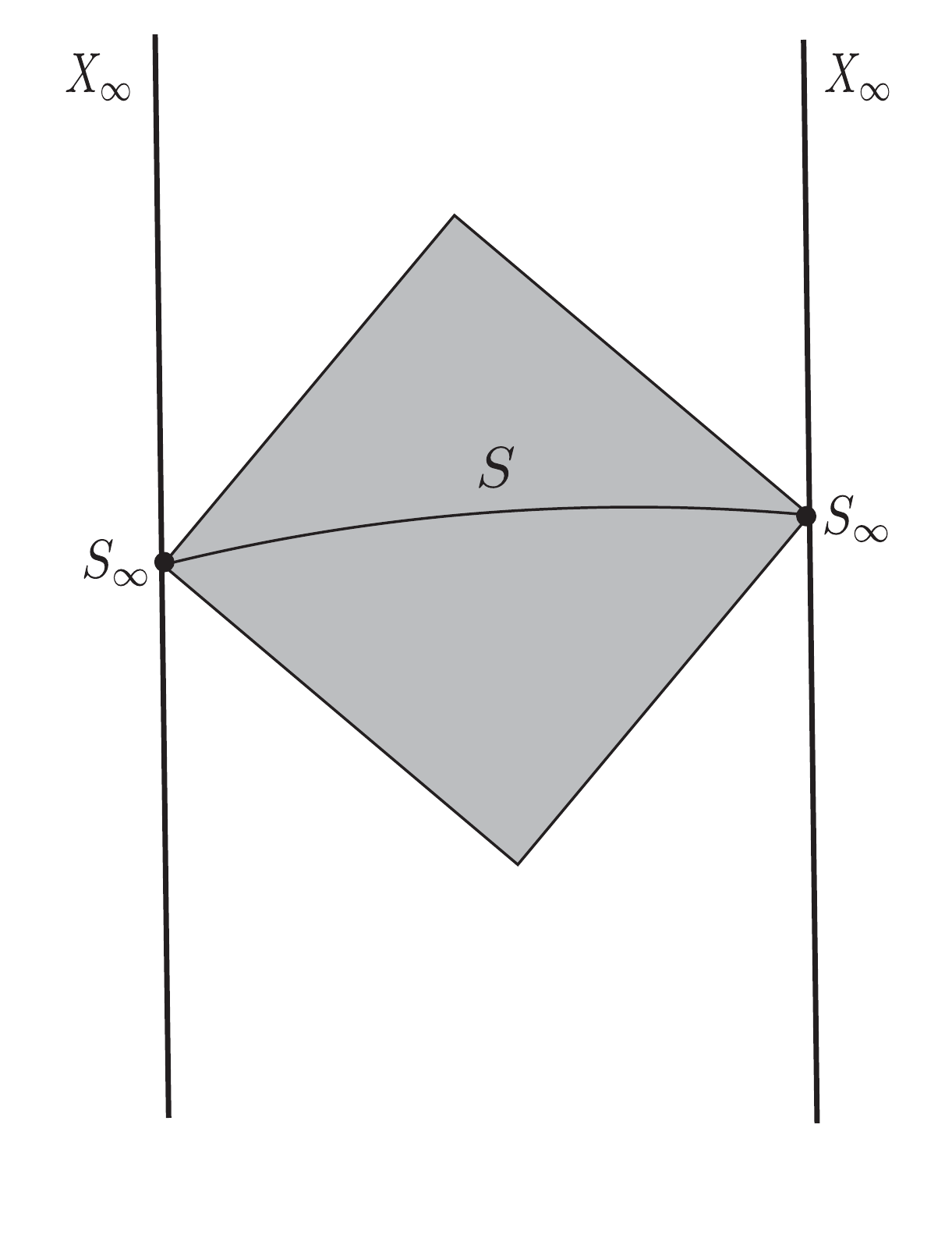}
 \end{center}
\caption{\small The bulk domain of dependence $\Omega$ of a Cauchy hypersurface $S_\infty$ in the boundary of an AAdS spacetime $X$.   In this
picture, for simplicity, $X$ is two-dimensional so its boundary $X_\infty$ is one-dimensional and $S_\infty$ consists of two points.   $\Omega$ is the domain
of dependence of any bulk Cauchy hypersurface $S$ with boundary $S_\infty$, or equivalently the set of bulk points that are not timelike separated from $S_\infty$.}
 \label{Domain}
\end{figure}

We will defer a detailed discussion of the classical phase space of asymptotically Anti de Sitter (AAdS) gravity to section \ref{class}.   Here we will explain a bare minimum to motivate an approach to the problem of describing a quantum Hilbert space.

The classical phase space of AAdS gravity is well understood in the case of pure gravity in three dimensions.    Let $X$ be an AAdS three-manifold, globally
hyperbolic in the AAdS sense,
that satisfies Einstein's equations with negative cosmological constant.   Its conformal boundary $X_\infty$ consists of one or more copies of $\R\times S^1$
(where the $\R$ and $S^1$ directions are respectively timelike and spacelike).   Let $S_\infty$ be any Cauchy hypersurface in $X_\infty$.   There are many
possible choices of bulk Cauchy hypersurface $S$ with boundary $S_\infty$, all homotopic to each other.\footnote{In the definition of $S$, we include the  conformal boundary
points in $S_\infty$.   This makes $S$ compact and 
generally enables simpler statements. Similarly, $S_\infty$ is included in the definition of the bulk domain of dependence $\Omega$.}   The bulk domain of dependence of $S_\infty$, which we will call $\Omega$, can be defined as the domain of dependence of
any $S$.   An alternative definition of $\Omega$ which makes it obvious that $\Omega$ does not depend on the choice of $S$ is that $\Omega$ consists of points in $X$
that are not timelike separated from $S_\infty$    (fig. \ref{Domain}).  
Thus $\Omega$ is a pseudo-Riemannian manifold with boundary.     $\Omega$ is the part of the spacetime that can be constructed,
given initial data on $S$, just using Einstein's equations, without using the AAdS boundary conditions along $X_\infty$.   

By the phase space $\vP$ of AAdS gravity in this situation, we mean the space of possible geometries  of the bulk domain of dependence $\Omega$, for a given
choice of $S_\infty$. 
 It is known (in the three-dimensional case)  that $\vP$ is actually a cotangent bundle,\footnote{This is also true in the case of a closed universe with $\Lambda<0$,
 though in this article, we mainly consider AAdS spacetimes.  In both cases, the same phase space $\vP$ has another description as a product of two copies
 of Teichm\"{u}ller space \cite{Mess}.   This description, which is suggested by the relation of three-dimensional gravity to Chern-Simons theory \cite{AT,Witten}, does not 
 generalize above three dimensions or in the presence of matter fields, so it is less relevant for a general understanding of gravity.}
  $\vP=T^*(\Conf/\Diff)$, where $\Conf$ parametrizes conformal structures  on $S$ (metrics modulo  Weyl transformations $h\to e^{2\varphi}h$ with $\varphi|_{S_\infty}=0$)
and $\Diff$ is the group of diffeomorphisms of $S$ that are trivial
 along $S_\infty$. Thus $\Conf/\Diff$ is the space of metrics on $S$ up to diffeomorphism and Weyl transformation.  That $\vP=T^*(\Conf/\Diff)$  is proved as follows  \cite{Monc,KS,BoS,SK}.  
 
  In one direction,  one makes use of the renormalized volume of a  hypersurface.    In an AAdS spacetime, a Cauchy hypersurface $S$ has infinite
  volume, but it is possible to define a renormalized volume $V_R(S)$.
  In three dimensions, one shows that, for any given $S_\infty\subset X_\infty$ and any choice of the bulk spacetime $X$,
  there exists a unique bulk
Cauchy hypersurface $S\subset X$ with boundary $S_\infty$ that maximizes $V_R(S)$.   (See section \ref{extremal} for a qualitative discussion of this existence and
uniqueness.) 
 $S$ has a Riemannian metric $h$ and a second fundamental form $K$;
extremality of $V_R(S)$ implies that $K$ is traceless, $\K=0$ where $\K=h^{ij}K_{ij}$. 
 Now, ``forget'' the metric $h$ and remember only the associated conformal structure,
which we will call $h_0$ (thus, knowing $h_0$ means knowing $h$ up to a Weyl transformation $h\to e^{2\varphi} h$).   
Then $h_0$, up to diffeomorphism, defines a point in $\Conf/\Diff$.   On the other hand, in General Relativity,   $K$ is canonically conjugate to $h$.
To be precise, the momentum conjugate to $h_{ij}$ is
\be\label{momcon}\Pi^{ij}=\frac{1}{16\pi G}\sqrt{h}\left(K^{ij}-h^{ij} K^r_r\right).\ee
The traceless part of this equation shows that 
the traceless part of $K$ is conjugate to the conformal structure $h_0$ (the trace  $\K=K^r_r$ is conjugate to the volume density $\sqrt{\det h}$, which
we abbreviate as $\sqrt h$).  
So the pair $K,h_0$, with $K$ being traceless, defines a point in $T^*\Conf$. 
To take diffeomorphisms into account, we have to divide by the group $\Diff$, but we also have to set to zero the Hamiltonian function on $T^*\Diff$ that
generates the action of $\Diff$.      Dividing by $\Diff$ removes from the phase space some modes of $h$ and setting the Hamiltonian function to zero
removes the conjugate modes of $K$. 
For $K$ traceless, the Hamiltonian function that generates $\Diff$ is $D_i K^{ij}$.
Setting the Hamiltonian function to zero is the momentum constraint of General Relativity.  These matters are explained  in sections \ref{constraint} and \ref{symplectic}.
  The combined operation of setting the Hamiltonian function to zero and dividing by $\Diff$ replaces $T^*\Conf$ with $T^*(\Conf/\Diff)$.
  So if $K$ and $h_0$ come from a solution of Einstein's equations, they define a point in $T^*(\Conf/\Diff)$.
The map that associates the pair $K,h_0$ to a given solution of the Einstein field 
equations therefore  gives a map from
the phase space $\vP$ to $T^*(\Conf/\Diff)$.     

To get a map in the opposite direction, one shows that given a point in $T^*(\Conf/\Diff)$, that is, a 
pair $K,h_0$ with $D_i K^{ij}=0$, one can in a unique fashion make a Weyl 
transformation to a pair that satisfies the Einstein constraint equations and thereby gives initial conditions for a solution of the full Einstein  equations, defining a point in $\vP$.
The proof is explained in detail in section \ref{class}.   The two maps are inverses, so the gravitational phase space can be identified as $\vP=T^*(\Conf/\Diff)$.

Are such ideas  relevant to a general understanding of gravity?   For this,  something similar should be true in higher dimensions, and also in the presence of
matter fields.    The full story of what is known to be true and what is likely to be true under reasonable assumptions is somewhat involved, and is deferred to 
section \ref{class}.   For now, we just remark that in the context of perturbation theory, in General Relativity on $\AdS_D$ or on $\AdS_D\times W$ 
for some compact manifold $W$,  possibly with matter fields, 
an analysis similar to what was just sketched  is always applicable. At least for purposes of perturbation theory, the phase space of such a theory can  always be represented by
a cotangent bundle $T^*\Q$, where now  $\Q$ parametrizes the conformal structure of $S$ together
with the matter fields on $S$, modulo diffeomorphisms that are trivial at infinity
(along with gauge transformations that are trivial at infinity if some of the matter fields are gauge fields).     That is true because
both steps in the construction -- existence of a unique $S$ of maximal volume, and existence of a unique Weyl transformation that ensures that the constraint equation
is satisfied
-- are  valid if one is sufficiently close to $\AdS_D$ or $\AdS_D\times W$.  As we will discuss in section \ref{class}, to extend these results beyond
perturbation theory, one requires a strong energy condition and a condition on singularities somewhat analogous to cosmic censorship.
But such assumptions are not necessary in perturbation theory around $\AdS_D$ or $\AdS_D\times W$.
Likewise, as we will also see in section \ref{class}, beyond perturbation theory
one can make much stronger statements for $\AdS_D$ than for $\AdS_D\times W$, but the difference is not relevant in perturbation theory.

\subsection{The Constraint Equations}\label{constraint}

The Einstein constraint equations are equations for a metric $h_{ij}$ on an initial value surface $S$ and a symmetric tensor field $K_{ij}$ on $S$.  These equations
are the condition under which $h$ and $K$ are initial data for a spacetime $X$ that satisfies Einstein's equations, with $h$ (whose scalar curvature will be denoted
$R(h)$) and $K$ understood as the induced metric and second fundamental form of $S\subset X$.   For General Relativity with cosmological constant $\Lambda$ and no matter fields, the equations read
\be\label{dofox}\P^j(\x)=\H(\x)=0,\ee
with 
\begin{align}\label{Zoneq}\P^j(\x)&=  D_i K^{ij}-D^j K^i_i   \cr
                          \H(\x)&=  R(h)-K^{ij} K_{ij}+K^i_i K^j_j-2\Lambda. \end{align}
                            The  equation $\P^j(\x)=0$ is called the momentum constraint and the equation $\H(\x)=0$ is called the 
                            Hamiltonian constraint.  Importantly, these are gauge 
                            constraints, which must be satisfied independently at each point $\x\in S$.
                            Quantum mechanically, $K$ is conjugate to $h$ as in eqn. (\ref{momcon}), so $\P^j(\x)$ and $\H(\x)$, for each $\x\in S$,
                            become differential operators acting on the space $\Met$ of metrics on $S$.   $\P^i(\x)$ is linear in $K$ so it becomes a first order differential operator which is simply  the generator   of diffeomorphisms
 of $S$;  to be precise, if $v^i$ is a vector field on $S$ then the generator of the symmetry\footnote{The symbol  $\bigdelta$ will denote a symmetry generator or the
 variation of a field,
 while $\delta$ will represent a Kronecker delta or the Dirac delta function.}  $\bigdelta x^i=v^i$         is $\int_S\d^{D-1}\x \sqrt{h} v^i(\x) \P_i(\x)$.
 $\H(\x)$ is quadratic in $K$, so it becomes a second order differential operator.

In the  most basic version of the canonical approach to quantum gravity, the quantum wavefunction is a function $\Psi(h)$ of the metric of $S$ (and possibly other
variables).
 The traditional interpretation of the constraint is that the operators obtained by quantizing the constraints should  annihilate $\Psi(h)$:
  \be\label{ancon}\P^i(\x) \Psi(h)=\H(x) \Psi(h)=0. \ee     
 Since $\P^i(\x)$ is the generator of diffeomorphisms, the constraint $\P^i(\x)\Psi=0$ merely says that $\Psi(h)$ should be invariant under diffeomorphisms of $S$ 
 (or more precisely, under those diffeomorphisms
 that are connected to the identity; it is generally assumed that this condition should be extended to all diffeomorphisms).    With this constraint imposed, $\Psi(h)$
 becomes a function on the space $\Met/\Diff$ of metrics modulo diffeomorphisms.   The Hamiltonian constraint $\H(\x)\Psi(h)=0$ is more vexing and more difficult 
 to interpret.  Because $\H(\x)$ is a second order differential operator for each $\x$, the 
  constraint $\H(\x)\Psi(h)=0$ is an infinite system of second order differential equations  that should be satisfied by the quantum 
  wavefunction.      This infinite system of equations (or sometimes the combined system $\P^i(\x)\Psi=\H(\x)\Psi=0$) is known as the Wheeler-DeWitt equation.   
 In the traditional approach to the canonical theory of gravity, a quantum wavefunction is a function on $\Met/\Diff$ that satisfies the Hamiltonian constraint equation, or
 equivalently a function on $\Met$ that satisfies the combined system $\P^i(\x)\Psi=\H(\x)\Psi=0$.
  
A basic difficulty of canonical quantum gravity is that it is very difficult to solve the Wheeler-DeWitt equation, or to gain any qualitative understanding of the solutions.
   However,  the fact that the phase space of General Relativity is a cotangent bundle $\vP=T^*(\Conf/\Diff)$ suggests a simple answer.   In general, a
 cotangent bundle $T^*Y$ can be quantized by saying
 that a physical state is a square integrable function\footnote{More canonically, since $Y$ may not have a natural measure, the wavefunction should be a half-density
 on $Y$ rather than a function on $Y$.  To avoid an inessential distraction, we will not always make this distinction.}  on the base space $Y$.  So
 one is led to hope that the state space of General Relativity can be interpreted as a space of functions on $\Conf/\Diff$, with no constraints.

The original suggestion along these lines was actually made by York half a century ago \cite{York}, for somewhat similar reasons to what was just
explained, and motivated by even earlier results that pointed in this direction.
For example,\footnote{The background to York's proposal also included parts of the story that will be described in section \ref{class}.}
Kuchar had shown \cite{Kuchar} that in asymptotically flat spacetime, to lowest nontrivial order, a solution $\Psi(h)$ of the Wheeler-DeWitt equation depends only on the 
transverse traceless part of $h$.  To be precise, here we perturb around the case that $S$ is a flat hypersurface $\R^{D-1}$ in $D$-dimensional Minkowski space
$\R^{D-1,1}$.   The metric of $S$ is thus taken to be $h_{ij}=\delta_{ij}+h'_{ij}$, where $\delta_{ij}$ is the Euclidean metric on $\R^{D-1}$ and $h'_{ij}$ is the perturbation.
Kuchar showed that to first order in $h'$, the Wheeler-DeWitt equations assert that the quantum wavefunction $\Psi(h)$ is completely determined by an 
arbitrary function of the transverse traceless part of $h'$.    In lowest order, the space of transverse traceless
metric perturbations is the same as the space of deformations of the conformal structure up to diffeomorphism, so Kuchar's result can be restated by saying that
to first non-trivial order, solutions of the Wheeler-DeWitt equation on $\Met/\Diff$ are in natural correspondence with functions on $\Conf/\Diff$.    
These arguments were recently
reworked in the AAdS case, with a similar result \cite{CGPR}.    

The relation of the Wheeler-DeWitt equation to the $T\bar T$ deformation and its generalizations \cite{Verl,KLM,Taylor,HKST,Shyam, ALS,AKW} actually gives
a way to generalize such statements to all orders in perturbation theory.   In explaining this, we will just consider the original example
 \cite{Verl}  of three-dimensional
pure gravity with $\Lambda<0$ and the original $T\bar T$ deformation \cite{Zam,ZamTwo,Cardy}.   The Wheeler-DeWitt equation of three-dimensional pure gravity
reads
\be\label{olfgo}\left(K^{ij}K_{ij}-K^i_i K^j_j -R(h)+2\Lambda\right)\Psi(h) =0 .\ee
Setting $\Lambda=-1/\ell^2$ and using eqn. (\ref{momcon}) to express $K$ in terms of $\Pi^{ij}(x)=-\i \frac{\delta}{\delta h_{ij}(x)}$, the equation becomes
\be\label{nolfgox} \left( \frac{(16\pi G)^2}{\det h} \left(\Pi^{ij}\Pi_{ij}-(\Pi^k_k)^2\right)-R(h)-\frac{2}{\ell^2}\right)\Psi(h)=0. \ee
Now  conjugate the constraint operator by $\exp\left(\frac{1}{8\pi G\ell} \int_S \d^2x \sqrt h\right) $ or equivalently  define
\be\label{nodigo}\Psi = \exp\left(\frac{1}{8\pi G\ell} \int_S \d^2x \sqrt h\right) \h\Psi.\ee
The effect of this change of variables is to shift    $\Pi^{ij}\to \Pi^{ij}-\frac{\i}{16\pi G\ell}\sqrt{ h}  h^{ij}$.
The equation satisfied by the new wavefunction $\h\Psi$ is\footnote{Here one throws away some terms formally proportional to $\delta(0)$ that come
from $\frac{\delta^2}{\delta h(x)^2}\sqrt{\det  h(y})\sim \delta(x,y)^2=\delta(x,y) \delta(0)$.  One can think of this step as a normal-ordering recipe. 
The $\delta(0)$ terms are subleading in $G/\ell$.} 
\be\label{onokey} \left(\frac{\i}{\sqrt{\det h}} \Pi^k_k + \frac{8\pi G\ell}{\det h} \left(\Pi^{ij} \Pi_{ij} -(\Pi^k_k)^2\right)-\frac{\ell}{32\pi G}R(h)\right)\h\Psi=0. \ee
The term $\frac{8\Pi G\ell}{\det h}\left(\Pi^{ij}\Pi_{ij}-(\Pi^k_k)^2\right)$ is irrelevant in the renormalization group sense; by power counting, it is negligible at long distances.
The leading long distance approximation to the equation is therefore simply
\be\label{tonokey} \left(\frac{\i}{\sqrt {\det h}} \Pi^k_k -\frac{\ell}{32\pi G}R(h)\right) \h\Psi(h)=0. \ee
This equation  is familiar in two-dimensional conformal field theory (CFT).  The operator $\i\Pi^k_k$ is the generator of Weyl transformations of the metric, and so 
eqn. (\ref{tonokey}) describes violation of conformal invariance by the usual $c$-number anomaly proportional to $R(h)$.  In fact, eqn. (\ref{tonokey}) is the usual
anomalous Ward identity of a two-dimensional CFT with central charge
\be\label{nokey} c=\frac{3\ell}{2G}, \ee
which is the  Brown-Henneaux formula for the central charge of the boundary stress tensor in three-dimensional gravity \cite{BH}. 
Eqn. (\ref{onokey}) differs from the usual CFT Ward identity by the $\Pi^2$ terms.  Since $\Pi^{ij}=-\i \frac{\delta}{\delta h_{ij}}$ inserts a factor of the stress tensor $T^{ij}$
in an amplitude of the boundary CFT, eqn. (\ref{onokey}) actually describes the combined violation of conformal invariance by the CFT anomaly along with
a $T\bar T$ deformation.   This was the main observation in \cite{Verl}.   If we factor the metric $h$ as $h=e^{2\varphi}h_0$, with some fixed choice of $h_0$,\footnote{We will learn in section \ref{class}
 that each Weyl orbit of metrics has a unique representative that satisfies the Hamiltonian constraint equation, so one could make this factorization by choosing $h_0$
 to be that representative.   A much more elementary way is to pick a smooth measure $\mu$ on $S$ and require $\sqrt{\det h_0}=\mu$.} 
 then we get $\Pi^i_i  =-\frac{\i}{2}\frac{\delta}{\delta\varphi}$ and
 eqn. (\ref{tonokey}) becomes the usual CFT Ward identity
  that determines the dependence of $\h\Psi$ on $\varphi$.
   In eqn. (\ref{onokey}), the $\Pi^2$ terms are of  relative order $e^{-2\varphi}$ compared to the other terms.
   So for large $\varphi$, eqn. (\ref{onokey}) reduces to the usual CFT Ward identity (\ref{tonokey}).
Any solution $\h\Psi_0$ of that CFT Ward identity can be promoted as follows to a solution $\h\Psi$ of the $T\bar T$-deformed equation (\ref{onokey}).
Let us denote the operators on the left hand sides of eqns. (\ref{onokey}) and (\ref{tonokey}) as $\mathcal D$ and $\mathcal D_0$, respectively.
We expand $\h\Psi = \h\Psi_0+\h\Psi_1 +\h\Psi_2+\cdots, $ and stipulate that for large $\varphi$, each $\h\Psi_k$ is of order $e^{-2k\varphi}$ relative to $\h\Psi_0$
and that
\begin{align}\label{huddo}{\mathcal D}_0\h\Psi_k=-{\mathcal D}\left(\h\Psi_0+\h\Psi_1+\cdots +\h\Psi_{k-1}\right)+\O(e^{-2(k+1)\varphi}\Psi_0). \end{align}
Order by order in $e^{-2\varphi}$, $\h\Psi_k$ is uniquely determined and eqn. (\ref{onokey}) is satisfied.   The expansion in powers of $e^{-2\varphi}$ is
equivalently an expansion in powers of $G$.   Thus, order by order in perturbation theory in $G$, $\h\Psi$ is uniquely determined in terms of $\h\Psi_0$.  Every
$\h\Psi$ arises in this way from some $\h\Psi_0$ (which can be found from the large $\varphi$ behavior of $\h\Psi$). Since the usual CFT  Ward identity determines the dependence of $\h\Psi_0$ on $\varphi$, this means that order by order in perturbation theory,
solutions $\h\Psi$ of the Wheeler-DeWitt equation are in natural correspondence with wavefunctions that depend on $h_0$ only, or in other words, functions on
$\Conf/\Diff$.   One can view this as a generalization of the result of \cite{Kuchar,CGPR} to all orders, in AAdS spacetime.

Knowing this correspondence does not immediate tell us the
correct form of the Hilbert space inner product on the space of solutions of the Wheeler-DeWitt equation.   
One can formally define a natural inner product for functions\footnote{Rather than functions on $\Conf/\Diff$, 
it is more natural to use half-densities, and really one  needs a more precise language that takes account of the conformal
anomaly.    We will not go in that direction, because such issues will not arise in the approach to constructing a Hilbert space that we
actually follow in section \ref{gaugefixing}.}   on $\Conf/\Diff$
by $(\Psi',\Psi)=\int_{\Conf/\Diff} \bar\Psi' \Psi$.    A more general inner product would be
\be\label{zelbo} \la\Psi',\Psi\ra =(\Psi'|\Xi|\Psi), \ee
for some positive self-adjoint operator $\Xi$.    In section \ref{gaugefixing}, we will show how a simple gauge-fixing leads to a description of the  perturbative Hilbert space
$\HH$ of quantum gravity, 
 roughly along these lines (but in a BRST formulation with ghost fields included), with a relatively simple and relatively explicit formula for $\Xi$ 
 as a ghost determinant.  
The derivation will also lead directly to formulas such as eqns. (\ref{formula}) and (\ref{formula2}), with  transition amplitudes expressed in terms of sums
over contributions of  intermediate states in $\HH$.     Such formulas are after all the goal of having a Hilbert space of physical states.
However, first we will say more in section \ref{wdw} about old and new approaches to the Wheeler-DeWitt equation
and how the procedure in section \ref{gaugefixing}  relates to them.

\subsection{The Wheeler-DeWitt Equation And The BRST Operator}\label{wdw}

The traditional interpretation of the Wheeler-DeWitt equation, going back to its origins, was as described in section \ref{constraint}: a quantum state
was taken to be a function $\Psi(h)$ of a $(D-1)$-geometry $h$, satisfying $\P^i(\vec x) \Psi(h)=\H(\vec x)\Psi(h)=0$.     

At least at a formal level, there is  a specific problem in which a wavefunction of this type actually arises.  

 \begin{figure}
 \begin{center}
   \includegraphics[width=4in]{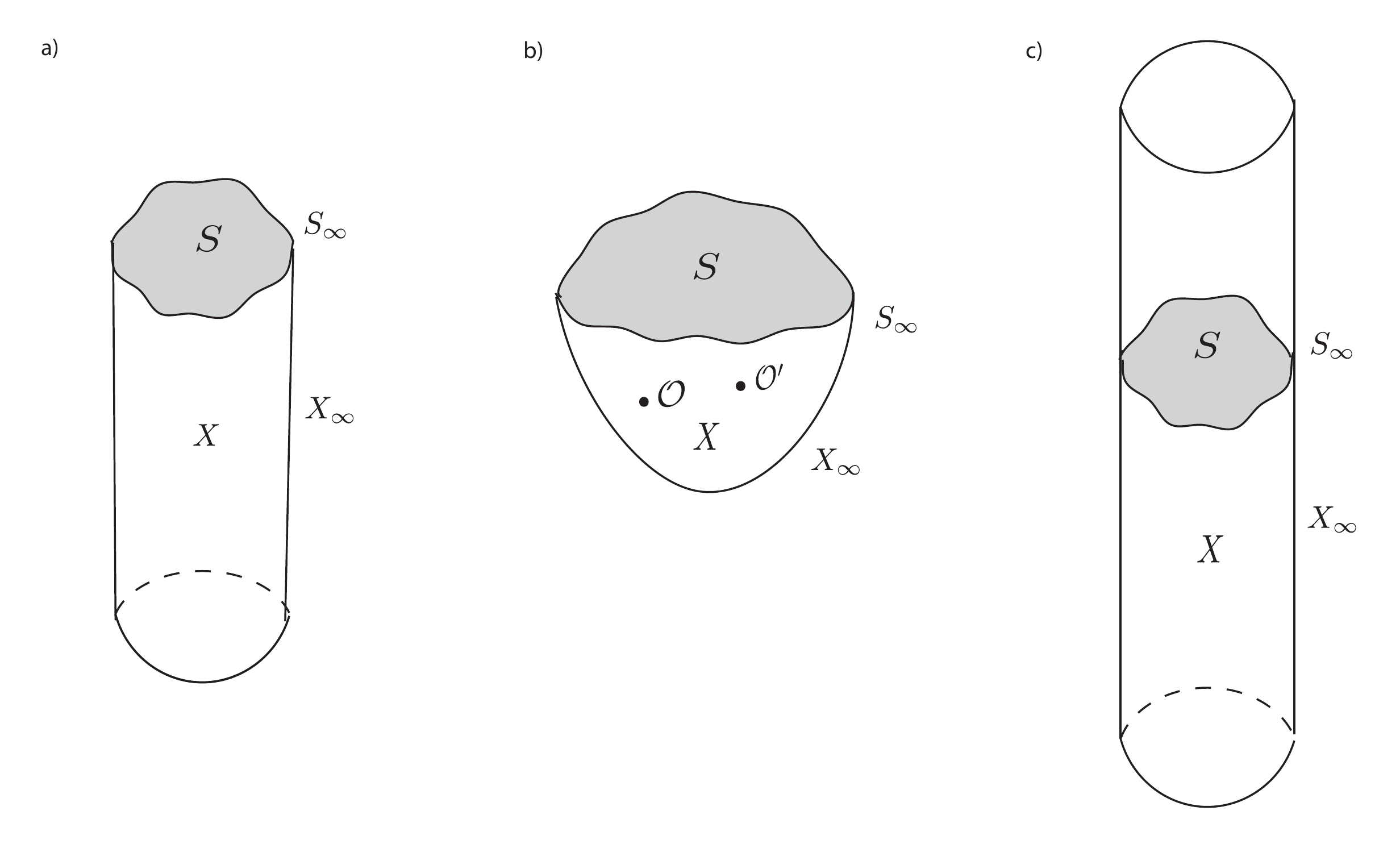}
 \end{center}
\caption{\small (a) Here  $X_\infty$ is a Lorentz signature boundary manifold with a future boundary $S_\infty$.   In the boundary theory, initial conditions in the far past,
and possible boundary insertions,  determine a quantum state  $\Psi_\infty$ on $S_\infty$.  The bulk is an AAdS manifold $X$ with future boundary $S$.  The metric
$h$ of $S$ is fixed and the path integral on $X$ defines a function $\Psi(h)$ which one hopes has the same physical content as $\Psi_\infty$.   One can
argue formally that $\Psi(h)$ satisfies the Wheeler-DeWitt equation. (b)  A similar picture to (a) in Euclidean signature.  The main difference is
that $X_\infty$ has operator insertions but no past boundary.   (c)  The picture of (a) is continued into the future and some final state is specified.
 In the boundary one gets nice formulas for the transition amplitude between specified initial and final states involving a sum over
states on $S_\infty$, but in bulk, there is a problem if the states are supposed to be solutions $\Psi(h)$ of the Wheeler-DeWitt equation.
If one picks a particular bulk Cauchy hypersurface $S$ on which to cut, the Wheeler-DeWitt equation is not satisfied, but integrating over all $S$ gives
a massive overcounting.}
 \label{AAA}
\end{figure} 

We consider some sort of initial conditions that, physically, should suffice to create a specific quantum state.   For example, in the AAdS context,
in Lorentz signature, we can do the following.  From a boundary point of view  (fig.  \ref{AAA}(a)), we specify a Lorentz signature manifold $X_\infty$ of dimension $D-1$
that starts at time $t=-\infty$ in the past and has a spacelike
future boundary $S_\infty$.   The boundary theory on $X_\infty$, with initial condition at $t=-\infty$ corresponding to some chosen state, and specified operator
insertions to the past of $S_\infty$,   will produce a quantum state  $\Psi_\infty$ on $S_\infty$. 
  One can also make a similar construction in Euclidean signature (fig. \ref{AAA}(b)).

To recover the state $\Psi_\infty$  from the gravitational  path integral, one considers a bulk
spacetime $X$ with conformal boundary $X_\infty$ at spatial infinity, and terminating in the future on a spacelike hypersurface $S$ whose boundary is
$S_\infty$.  
 We write $g$ for the metric of $X$ and $h=g|_S$ for the induced metric
on $S$.     Now we do a bulk path integral, with  initial conditions and boundary insertions as before, and with Dirichlet boundary conditions keeping fixed the metric $h$ of $S$.
The output of the path integral is a function $\Psi(h)$.
  This function is
supposed to define the state on $S$ created by the gravitational path integral under the given conditions.

The main virtue of this construction is that one can argue formally that $\Psi(h)$ satisfies the Wheeler-DeWitt equation.   As the construction is manifestly
invariant under diffeomorphisms of $S$, it is evident that $\P^i(\x)\Psi(h)=0$, and one can  argue formally that $\H(\x)\Psi(h)=0$ (see for example
\cite{HH,Hall,Barv, Matz}). One can conjecture that $\Psi(h)$ is a bulk
dual of the boundary state $\Psi_\infty$.

This construction  has two drawbacks.   The first is that the problem that was solved is not really the problem that one wanted to solve.
The reason for wanting  to construct a Hilbert space of quantum states is that one wants to be able to factorize transition amplitudes in terms of sums over
intermediate states, as in eqns. (\ref{formula}) and (\ref{formula2}).   After all, this is what quantum states are good for in ordinary quantum mechanics.  Although
one can argue at a formal level that the wavefunction $\Psi(h)$ created by the gravitational path integral satisfies the Wheeler-DeWitt equation, there is no argument
even formally that such wavefunctions participate in the desired ``sum over states'' formulas.   The reason is that when 
we compute a path integral that we want to evaluate by summing over intermediate states, there is no natural way in the context 
of Dirichlet boundary conditions to find the bulk hypersurface $S$ whose metric $h$ we should be summing over (fig. \ref{AAA}(c)).   General
covariance would force us to integrate over all choices of $S$, which involves a massive overcounting.   

The second drawback is that the gravitational path integral that is supposed to compute $\Psi(h)$ is actually not well-defined even in perturbation theory (and even
after regularizing ultraviolet divergences), because the Dirichlet boundary condition that was assumed is not elliptic    \cite{AE,Anderson,WittenBC}.  This lack
of ellipticity means
that, with Dirichlet boundary conditions, the operator $L$ that arises by linearizing the gauge-fixed Einstein equations about a classical solution does not have a 
well-defined determinant or propagator.\footnote{An exception is that
if, classically, the universe is everywhere expanding or everywhere contracting along the boundary (and more generally if the canonical momentum
is everywhere a  positive- or negative-definite matrix along the boundary), the determinant and propagator
may be well-defined even
though the boundary condition is not elliptic.   This is explained in \cite{WittenBC}, following \cite{AndersonTwo}.}

One might be inclined to dismiss the second problem as a technicality.   However, if one actually tries to actually compute 
$\Psi(h)$ in perturbation theory in a specific situation, one will soon need
the determinant and propagator of the operator $L$, and one will run into difficulties.   There is actually another reason to believe that the non-ellipticity of
the Dirichlet boundary condition
on $L$  should not be dismissed lightly.   This non-ellipticity can be straightforwardly proved, with a little linear
algebra, starting from the definition of an elliptic boundary condition \cite{AE,WittenBC}.   However, there is a more abstract proof that is quite instructive \cite{Anderson}.
In this  proof, the only real input is the form of the Hamiltonian constraint equation for gravity and specifically the fact that it involves second derivatives of $h$ along the
boundary  (which appear in the scalar curvature $R(h)$) 
but only first derivatives in the normal direction  to the boundary (which are present  in the definition of $\H(\vec x)$ in eqn. (\ref{Zoneq})  because $K$ is linear in the normal derivative of $h$).  The Hamiltonian constraint equation
is the cause of the difficulty in understanding the canonical formalism for gravity, so 
 in trying to understand that canonical formalism, we probably should not ignore a mathematical problem associated to the form of the constraint equation.
 
The problem involving the lack of ellipticity 
has a simple fix.   Instead of Dirichlet boundary conditions for gravity in which one fixes the boundary metric,\footnote{Neumann boundary condtions,
in which one fixes the second fundamental form $K$ of the boundary rather than the boundary metric $h$, are again not elliptic \cite{Anderson}.} one
can consider a mixed Dirichlet-Neumann boundary condition in which one specifies not the boundary metric $h$, but rather the conformal structure $h_0$ of
the boundary (in other words, the boundary metric up to a Weyl transformation) and the trace $\K=K^i_i$ of the second fundamental form $K_{ij}$.    
This mixed Dirichlet-Neumann boundary condition is elliptic \cite{Anderson,WittenBC}, 
so in the situation of fig. \ref{AAA}(a), it should be possible in perturbation theory, after regularizing ultraviolet divergences, to use this boundary condition
 to compute a wavefunction $\Psi(h_0,\K)$.  

One drawback of this  is that the Wheeler-DeWitt equation in a dual version with $\K$ treated as a coordinate and $\sqrt {h}$ as a conjugate momentum appears
to be, at best, no simpler than the original.   Another and possibly more serious problem is that, again, 
this construction seems to solve the wrong problem.   It formally gives a way to solve the problem described in fig. \ref{AAA}(a), but the problem of fig. \ref{AAA}(c)
remains.  There is no argument even formally 
that  a gravitational path integral can be evaluated by ``cutting'' on an intermediate
hypersurface $S$ and summing over states on $S$ of the form $\Psi(h_0,\K)$ that satisfy the constraint equations.

There is, however, also  a standard fix for this difficulty.   So far we have described what might be called
 the ``traditional'' Wheeler-DeWitt formalism.  There is also a ``revised'' Wheeler-DeWitt formalism
in which one constructs states that are better candidates for appearing in a factorization formula \cite{Nacht,Higuchi,Asht,MarolfOther,Embacher,GM, Marolf}  (see \cite{AKW} and 
Appendix B of \cite{CLPW}
for  recent discussions).   In the revised Wheeler-DeWitt formalism, sometimes called refined algebraic quantization or group averaging,
one still considers a wavefunction $\Psi(h)$, and one still imposes the momentum constraint equation $\P^i\Psi(h)=0$.    However, the constraint $\H\Psi=0$
is replaced by an equivalence relation 
\be\label{zondox} \Psi(h)\cong \Psi(h)+\sum_i \H(\x_i) \chi_i(h) \ee
for an arbitrary set of points $\x_i\in S$ and arbitrary functions $\chi_i(h)$.  (The discrete sum
over points $\x_i\in S$ can also be replaced by a continuous integral.)   In other words, the sense in which $\H(\x)$ vanishes
 is not that it annihilates a physical state, but that its action is trivial, since  
any state $\H(\x)\chi$ is considered trivial.    In  this approach, any state $\Psi(h)$ that satisfies the momentum constraint is considered physical;
two such states are considered equivalent if their difference is of the form $\sum_i \H( \x_i)\chi_i(h)$.

 \begin{figure}
 \begin{center}
   \includegraphics[width=3in]{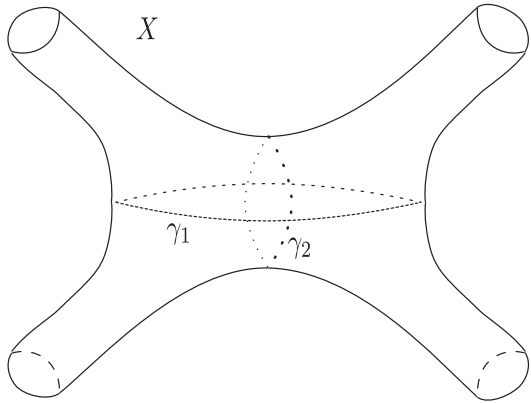}
 \end{center}
\caption{\small An AAdS spacetime $X$ with four asymptotic regions in which asymptotic states might be specified. Such an $X$ cannot have a metric
everywhere of Lorentz signature; it may have Euclidean signature or possibly a complex metric.  Overlapping ``cuts'' of such an $X$ can be made,
as sketched here,  on
homotopically inequivalent surfaces such as $\gamma_1$ and $\gamma_2$.  No one canonical formalism is well adapted to all of the possible cuts.}
 \label{BBB}
\end{figure} 

In this revised Wheeler-DeWitt approach, the inner product of two states is defined formally as
\be\label{ondo}\la \Psi'|\Psi\ra=\int _{\Met/\Diff}Dh\,\,  \b\Psi'(h)\prod_{\x\in S}\delta(\H(\x))\Psi(h) .\ee
Here $Dh$ represents formally an integral over the space $\Met/\Diff$ of metrics on $S$ up to diffeomorphism.   The product of delta functions
$\prod_{\x\in S}\delta(\H(\x))$ formally annihilates any state of the form $\H(\x)\Chi$, ensuring invariance of the inner product under the equivalence relation.

With this revised interpretation of the constraint operators, it is possible to give a formal argument that leads to the desired formulas involving cutting and
summing over intermediate states, as in eqn. (\ref{formula}).   For this, one goes to a canonical ADM formulation of the path integral in the region in which cutting is
supposed to happen.   In that formulation, the action contains a term $\int_S\d^{D-1}\x\,\, N(\x) \H(\x)$, where $N$ is called the lapse and does 
not appear elsewhere in the action.   The path integral therefore contains a factor
\be\label{zolto}\int DN \exp\left(\i \int_S\d^{D-1}\x\,N(\x)\H(\x)\right). \ee
Assuming that $N(\x)$ is supposed to be integrated from $-\infty$ to $+\infty$, the integral over $N$ gives formally the desired $\prod_{\x} \delta(\H(\x))$.

A possible criticism of this approach -- the status of this issue is not clear to the author  -- is that in replacing the covariant version of the Einstein path integral with a canonical version in which $N$ is allowed to
have either sign, we may have changed the path integral in a way that was adapted to the specific cutting formula we were trying to get.  In the covariant path
integral (or classically), it looks natural for $N$ to be positive.     We really want to know how to evaluate  the original covariant form of the path integral 
 by a cutting formula.   This issue is particularly sharp in a Euclidean 
signature  context in which the boundary theory may satisfy many different formulas that result from  cutting on topologically inequivalent 
hypersurfaces (fig. \ref{BBB}).   No one canonical version of
the bulk path integral can reproduce all of those different cutting formulas, so if one is going to use canonical versions of the path integral to 
deduce cutting formulas, it is essential to know that these canonical versions are all equivalent to the underlying covariant version of the path integral.

The traditional and revised Wheeler-DeWitt theories can be viewed as two special cases of what one can do with BRST quantization.   In BRST quantization, 
one introduces ghost fields, of ghost number 1, that transform like the generators of the gauge symmetries, but with opposite statistics.   In the case of gravity, 
the ghost
fields are an anticommuting  vector field $c^\mu(\vec x,t)$.    One also introduces additional multiplets consisting of antighost fields and auxiliary fields; this part of the construction
is nonuniversal and depends on what gauge condition one wishes to impose.  The BRST operator, in the context of gravity, is \cite{Frad}
\be\label{dorfo}Q=\int_S \d^{D-1}\vec x \sqrt{h}\left( c^0(\vec x) \H(\vec x) +c^i(\vec x) \P_i(\vec x)+\cdots \right) , \ee
where the omitted terms do not affect the following remarks.  This operator obeys $Q^2=0$, so one can define its cohomology.  As usual, the cohomology of $Q$ is defined to consist
of states $\Psi$ that satisfy $Q\Psi=0$, modulo the equivalence $\Psi\cong \Psi+Q\chi$ for any $\chi$.   In BRST quantization, the 
 cohomology  at one particular (theory-dependent) value of the ghost number is defined as the
Hilbert space of physical states.   In the case of gravity, if we assume that we are interested in states that are not annihilated by any modes of $c^0$ and $c^i$
(and that therefore are annihilated by all modes of the conjugate antighosts), the condition $Q\Psi=0$ gives $\H(\x)\Psi=\P_i(\x)\Psi=0$, the traditional
Wheeler-DeWitt constraints.   On the other hand, we could assume that we are interested in states that are annihilated by all modes of $c^0$ but not by any
modes of $c^i$.    Then the condition $Q\Psi=0$ gives the momentum constraint $\P_i(\vec x)\Psi=0$, but not the Hamiltonian constraint $\H(\vec x)\Psi=0$.
Instead, the equivalence $\Psi\cong \Psi+Q\chi$ leads to the equivalence relation (\ref{ondo}) of the revised Wheeler-DeWitt approach.

Thus the traditional and revised Wheeler-DeWitt theories are special cases of what one can do in the BRST framework.   Neither of these  corresponds 
closely to the  way that BRST quantization is usually carried out in ordinary gauge theory or in perturbative string theory.   Usually, the starting point is a relatively
standard Fock space of ghost and antighost fields, with a basis of states that are annihilated by roughly half of the ghost modes and half of the antighost
modes.   In other words, in setting up the BRST machinery  and using it to define the physical Hilbert space, ghosts and antighosts are usually treated rather similarly to other fields.

In the next section, we will describe a simple gauge-fixing that can be used to construct a Hilbert space for gravity.  The construction is valid to all orders of perturbation
theory, but not beyond, at least not in the present formulation.   A factorization formula is manifest.   The states that appear in 
the factorization formula are functions on $\Conf/\Diff$, the answer that is suggested
by the relation of the gravitational phase space to $T^*(\Conf/\Diff)$.   The boundary condition that is used in defining these states is the elliptic Dirichlet-Neumann
boundary condition.   The BRST approach to quantization is used, but not in the way that leads to either
the traditional or the revised Wheeler-DeWitt theory.  A fairly explicit formula for the inner product will emerge.

\subsection{A Simple Gauge-Fixing To Construct A Perturbative Hilbert Space}\label{gaugefixing}

The part of the BRST formalism for gravity that is universal involves the metric tensor $g_{\mu\nu}$ and the ghost field $c^\mu$.
They transform as
\be\label{howto}\largedelta g_{\mu\nu}=D_\mu c_\nu+D_\nu c_\mu, ~~~\largedelta c^\mu = c^\alpha \partial_\alpha c^\mu, \ee
where $\bigdelta$ represents the infinitesimal deformation generated by the BRST charge $Q$.   These formulas satisfy $\bigdelta^2=0$, which 
corresponds to $Q^2=0$; since $\bigdelta^2=0$, any expression of the general form $\bigdelta \Lambda$ is BRST-invariant, for any
 $\Lambda$.   
 
   The rest of the BRST formalism depends on what gauge-fixing condition one wishes to impose.
In general the desired gauge condition may be defined by a family of equations
\be\label{owlo} P_\Upsilon(g)=0, \ee
where we do not specify the nature of the labels $\Upsilon$ carried by these equations. (More generally, the $P_\Upsilon$ could depend on matter fields as well
as on the metric.)   To impose such a gauge condition, we add a family of
antighost fields $\bar c_\Upsilon$ and auxiliary fields $\phi_\Upsilon$ with
\be\label{towlo}\largedelta\bar c_\Upsilon=\phi_\Upsilon,~~~\largedelta\phi_\Upsilon=0, \ee consistent with $\bigdelta^2=0$.
 A simple way to implement a gauge-fixing that will impose the condition $P_\Upsilon(g)=0$ is
to add to the action a gauge-fixing term
\begin{align}\label{nowlo}I_\gf&=\largedelta\left(\sum_\Upsilon \bar c_\Upsilon P_\Upsilon \right)=\sum_\Upsilon \left(\phi_\Upsilon P_\Upsilon- \bar c_\Upsilon \largedelta P_\Upsilon\right)
\cr &=\sum_\Upsilon\left(\phi_\Upsilon P_\Upsilon -\bar c_\Upsilon \int_X \d^Dx \frac{\largedelta P_\Upsilon}{\largedelta g_{\mu\nu}(x)}(D_\mu c_\nu(x) + D_\nu c_\mu(x))\right). \end{align}
Thus, if we add to the action no other terms\footnote{In practice, it is often convenient to add to the action another term $-\frac{1}{2}\sum_\Upsilon\bigdelta(\bar c_\Upsilon
\phi_\Upsilon)=-\frac{1}{2}\sum_\Upsilon \phi_\Upsilon^2 +\dots $ (where the omitted terms involve fermions).   Then one can perform a Gaussian integral
over $\phi_\Upsilon$, leaving a contribution  $\frac{1}{2}\sum_\Upsilon P_\Upsilon^2$ to the action for the metric.   This can be more convenient than a delta function
constraint $P_\Upsilon=0$.} that involve $\phi_\Upsilon$, then $\phi_\Upsilon$ will behave as a Lagrange multiplier, imposing a gauge condition $P_\Upsilon=0$.

This procedure can be used to impose quantum mechanically any gauge condition that would be correct classically.   A gauge condition is correct classically if
on the diffeomorphism orbit of $g_{\mu\nu}$, there is a unique representative with $P_\Upsilon=0$.    In practice, one usually has to content oneself with a gauge
condition that is correct classically in the context of perturbation theory -- in other words, a gauge condition that is correct on gauge orbits that are sufficiently
close to some starting point.   For topological reasons, it is usually not possible to find a gauge condition that is uniformly valid on all gauge orbits.

In the case of gravity, assuming that one is constructing perturbation theory in an expansion around a classical solution $g_0$ of Einstein's equations,
one can write the full metric as $g=g_0+g_1$, and impose a gauge condition on the perturbation $g_1$.   
A simple and convenient gauge condition (which goes by names such as harmonic, de Donder, or Bianchi gauge) is to require $T^\mu(x)=0$ with
\be\label{zono} T^\mu(x)=D_\mu g_1^{\mu \nu}- \frac{1}{2}D^\nu g_{1\mu}^\mu, \ee
where covariant derivatives are taken with respect to the background metric $g_0$, and indices are also raised and lowered with that metric.  Thus with this choice,
the label $\Upsilon$ of the general discussion corresponds to a point $x\in X$ and an index $\mu$.

Here we will modify the gauge-fixing procedure so that it will help us in solving the problem identified in fig. \ref{AAA}(c).  Given a Cauchy hypersurface
$S_\infty$ in the boundary $X_\infty$ of an AAdS spacetime $X$, from a boundary point of view, a transition amplitude between initial and final states  can be factored as a sum over contributions of quantum states on $S_\infty$.   We want to obtain a similar description from a bulk point of view.  

If $X$ is actually $\AdS_D$ for some $D$, then it is shown in \cite{BoS} that any boundary  Cauchy hypersurface $S_\infty$ is the conformal boundary of a unique
bulk Cauchy hypersurface $S$ of maximal renormalized volume $V_R(S)$.   A similar result has been obtained much more recently
 \cite{CG} in a spacetime that is asymptotic to $\AdS_D$, provided the
bulk domain of dependence is compact.  The role of this assumption is explained in section \ref{exi}; for $D=3$, the assumption  is not necessary \cite{Monc,KS,BoS,SK}.   For a spacetime asymptotic to $\AdS_D\times W$ for some $W$,
as discussed in section \ref{exi}, we expect the maximal volume hypersurface $S$ to exist whenever the bulk domain of dependence is compact, but rigorous
results along these lines are not available at present.\footnote{The condition on the bulk domain of dependence is necessary; see section
\ref{compactification}.}
 However, to construct  perturbation theory,  one  does not need such strong results.  In perturbation theory, we expand around some sort of classical
 limit.    Typically this classical limit involves  a spacetime $X$ and a boundary Cauchy hypersurface $S_\infty$ such that the bulk Cauchy hypersurface $S$ of maximal volume does exist
and is unique.  For example, if $X=\AdS_D\times W$ for some $W$, then with a standard choice of $S_\infty$, the unique maximal volume hypersurface is  $S=\AdS_{D-1}\times W$, and we
can take this as the starting point of perturbation theory.      
In any such case,  the elliptic nature of the equation for a Cauchy hypersurface to have maximal volume ensures that after any sufficiently small perturbation
  of $X$ and/or $S_\infty$, a volume-maximizing $S$ that is asymptotic to $S_\infty$ still exists and is unique.   Under such conditions, this existence and uniqueness can be assumed
  to all orders of perturbation theory.

In perturbation theory, we integrate over different possible metrics on $X$, and until a metric is given, of course we do not know which hypersurface $S$ of
boundary $S_\infty$ is the Cauchy hypersurface of maximal $V_R(S)$.    However, we can proceed as follows.   Pick an arbitrary hypersurface $S_0\subset X$
with boundary $S_\infty$ that topologically is a potential Cauchy hypersurface.  Without loss of generality, we can pick a ``time'' coordinate $t$ on $X$
such that $S_0$ is defined by $t=0$.   (Unless a special choice was made of $S_\infty$, this coordinate $t$ does not restrict to anything standard on $X_\infty$.)   
Now suppose given an AAdS metric $g$ on $X$, sufficiently close to the standard one.   For this AAdS metric, there will be some bulk Cauchy hypersurface $S$
that maximizes $V_R(S)$.   Since $S_0$ is a potential Cauchy hypersurface and $S$ is another,  there is some diffeomorphism of $X$ that maps $S$ isomorphically onto $S_0$.   

This suggests the following strategy for gauge-fixing of quantum gravity on $X$.   As a first step in the gauge-fixing, we fix a small part of the diffeomorphism symmetry
by requiring $S=S_0$.   Then we perform gauge-fixing to the past and future of $S_0$ in any standard fashion, for instance via the harmonic gauge condition.
How one does that will not be important in what follows.   All that is important is that one of the gauge conditions is $S=S_0$.

The condition for a hypersurface $S_0$ with second fundamental form $K_{ij}$ to extremize the renormalized volume is $\K=0$, where $\K=K^i_i$ is the trace  of $K$.
(This standard fact will be verified shortly.) 
So the gauge condition that we want is that the surface $S_0$, which is defined by $t=0$, has $\K=0$.

To impose the gauge condition that $\K=0$ on the hypersurface $S_0$, we introduce a BRST multiplet consisting of a pair of scalar fields $b,\phi$ that are defined only
on that hypersurface, and satisfy the usual BRST transformation laws of antighost multiplets:
\be\label{ziflo}\largedelta b=\phi,~~~\largedelta \phi=0. \ee
Here $b$ is a fermion with ghost number $-1$ and $\phi$ is a boson with ghost number zero.   
We then introduce the partial gauge-fixing action
\be\label{miflo}\largedelta\int_S\d^d x  \sqrt h b \K =\int_S\d^d x  \sqrt h\left( \hat\phi \K -b \largedelta \K\right),\ee
with $\h\phi=\phi+(\bigdelta\sqrt h)b/\sqrt h$.
The field $\h\phi$ behaves as a Lagrange multiplier setting $\K=0$ on $S$.  

In eqn. (\ref{miflo}), $\bigdelta\K$ is the BRST variation   of $\K$ on the hypersurface $t=0$.    This BRST variation comes from the variation of the metric $g$ which
enters the definition of $\K$:   $\bigdelta g_{\mu\nu}=D_\mu c_\nu + D_\nu c_\mu$.     The ghost field $c^\mu$ has components $c^i$ associated to vector fields
that generate diffeomorphisms
of the $t=0$ hypersurface $S_0$, and a component $c^0$ that generates shifts of $t$.   Since the condition $\K=0$ is invariant under diffeomorphisms of $S_0$,
$\bigdelta\K$ is actually independent of $c^i$ if $\K=0$ and is $-\Xi c^0$ for some linear operator $-\Xi$.  So 
\be\label{belfo} -\int_S \d^dx\sqrt h\, b\largedelta\K =\int_S\d^d x\sqrt h \, b\Xi c^0  . \ee

A convenient way to identify $\Xi$ is as follows.
  We can pick local coordinates $t$ and $\vec x=x^1,\cdots , x^d$ near $S_0$ such that $S_0$ is defined by the
condition $t=0$, and the metric near $S_0$ has the form\footnote{One uses the orthogonal geodesics to the hypersurface $S_0$ to put the metric
locally in this form.   See for example section 4.3 of \cite{Wittenrays}.}
\be\label{metform}\d s^2 =-\d t^2 +\sum_{i,j=1}^d g_{ij}(\vec x,t) \d x^i \d x^j. \ee
 We expand $g$ around $\epsilon=0$ and write just $h$, $\dot h$, $\ddot h$ for the coefficients:
\be\label{etform} g(\vec x,\epsilon)=h(\vec x)+\epsilon \dot h(\vec x) +\frac{1}{2}\epsilon^2 \ddot h(\vec x) +\O(\epsilon^3). \ee
It is convenient to define the volume density  $v(x,\epsilon)=\sqrt{\det g(\vec x,\epsilon)}$. 
The second fundamental form of $S_0$ is 
\be\label{second} K_{ij}=\frac{1}{2} \dot h_{ij}\ee
  and its trace is
\be\label{traceform}\K=\frac{1}{2}h^{ij}\dot h_{ij}=\left.\frac{\dot v}{v}\right|_{\epsilon=0}. \ee

Now consider a general nearby Cauchy hypersurface $S$ defined by $t=\epsilon(\vec x)$.  To first order in $\epsilon$, its volume is just
\be\label{wondo}V(S)=\int_{S_0}\d^d x \sqrt{\det g(\vec x,\epsilon)}= V(S_0)+\int_{S_0}\d^d x \sqrt{h} \,\epsilon\K+\O(\epsilon^2). \ee
So the condition for $S_0$ to have extremal volume is $\K=0$.    We have written eqn. (\ref{wondo}) naively in terms of the ordinary volumes,
ignoring the fact that in the AAdS context, these volumes are divergent.  One actually wants to express formulas such as eqn. (\ref{wondo}) in terms of the
renormalized volume.  To define the renormalized volume $V_R(S)$, one restricts the integral over $S_0$ in the definition of the volume to a large compact region, and
then one subtracts some locally defined counterterms near the boundary and removes the cutoff.  If $\epsilon(\vec x)$ vanishes sufficiently rapidly at infinity, 
the counterterms are the same
for $S$ and $S_0$ and we can rewrite eqn. (\ref{wondo}) in terms of renormalized volumes:
\be\label{zondo}V_R(S) =V_R(S_0) +\int_{S_0} \d^d x\sqrt h\epsilon \K+\O(\epsilon^2). \ee

The shows that a necessary condition for $S_0$ to have maximal, or even extremal, renormalized volume is that it satisfies $\K=0$.   However, to identify the operator
$\Xi$, we need to compute $\K$ not for the hypersurface $S_0$, but for a nearby hypersurface $S$ with $t=\epsilon(\vec x)$.  Since we have learned that $\K$ is the
derivative of the renormalized volume with respect to $\epsilon$, one way to compute the $\O(\epsilon)$ contribution to $\K$ is to compute the renormalized volume
including terms of order $\epsilon^2$.   Differentiating the resulting formula with respect to $\epsilon$ will then give $\K$ including terms of first order.

A straightforward calculation gives the volume of $S$ including terms of order $\epsilon^2$:
\be\label{nefform}V_R(\S) =V_R(S_0) +\int_{S_0} \d^d x \sqrt{h}\left( \epsilon \frac{\dot v}{v} +\frac{\epsilon^2}{2} 
\left(\frac{\dot v}{v}\right)^2 +\frac{\epsilon^2}{2}\partial_t\left(\frac{\dot v}{v}\right)
-\frac{1}{2}g_0^{ij}\partial_i \epsilon \partial_j\epsilon \right) 
+\O(\epsilon^3) .\ee
To put this in a convenient form,  we use Raychaudhuri's equation for the $tt$ component of the Ricci tensor,\footnote{This is the
original timelike version of Raychaudhuri's equation \cite{raych}, not the null version \cite{sachs} that governs causal structures.}    which says that at $t=0$,
\be\label{inco}R_{tt}=-\partial_t\left(\frac{\dot v}{v}\right) -K_{ij}K^{ij}. \ee
Using also Einstein's equation $R_{tt}=8\pi G \h T_{tt}$, where $T_{\mu\nu}$ is the matter stress tensor (including a contribution from the cosmological constant) and
$\h T_{\mu\nu}=T_{\mu\nu}-\frac{1}{D-2}g_{\mu\nu}T^\alpha_\alpha$, we see that
if $S_0$ is an extremal surface, with $\dot v|_{t=0}=0$, then the renormalized volume of $ S$ to quadratic order in $\epsilon$ is
\be\label{winco}V_R(S)=V_R(S_0)-\int_{S_0}\d^dx \sqrt{h}
\left( \frac{1}{2}h^{ij}\partial_i \epsilon(\vec x) \partial_j \epsilon(\vec x) +\frac{1}{2}\epsilon(\vec x)^2 \left(8\pi G \h T_{tt}+K_{ij}K^{ij}\right)\right)+\O(\epsilon^3). \ee
Let $\Delta=h^{ij}D_i D_j$ be the Laplacian of the hypersurface $S_0$, acting on scalar fields.   Varying $V_R(S)$ with respect to $\epsilon$, we get
\be\label{zinco}\largedelta V_R(S)=-\int_{S_0}\ d^d x \sqrt h \,\largedelta\epsilon\left( -\Delta +8\pi G\h T_{tt} +K_{ij}K^{ij} \right) \epsilon. \ee 
Comparing to eqn. (\ref{zondo}), we can read off the term in $\K$ that is linear in $\epsilon$:
\be\label{wincox}\K=-\left(-\Delta +8\pi G\h T_{tt}+K^{ij}K_{ij}\right) \epsilon. \ee

For our application, we simply take $\epsilon$ to be the time component $c^0$ of the ghost field.   The infinitesimal diffeomorphism generated by this
field maps the hypersurface $t=0$ to the hypersurface $t=c^0$.  So the BRST variation $\bigdelta \K$ of $\K$ is obtained 
by substituting $\epsilon=c^0$ in eqn. (\ref{winco}).  Thus
\be\label{linco}\largedelta \K=-\Xi c^0,~~~~\Xi = -\Delta+8\pi G\h T_{tt}+K^{ij}K_{ij}, \ee
and the action (\ref{miflo}) associated to the partial gauge-fixing that makes  $S_0$ the maximal volume hypersurface is
\be\label{polinco} \int_{S_0}\d^d x \sqrt h\left(\h\phi\K +b\Xi c^0\right). \ee 
In putting the gauge-fixing action in this form, we made use of Einstein's equations for $R_{tt}$.  Quantum mechanically, this means that a field redefinition is involved
in putting the gauge-fixing action in this form.   

The partial gauge-fixing condition that we have used is only satisfactory if the operator $\Xi$ has no zero-mode. Otherwise, there is a mode of $b$ that
decouples from the action,  the path integral will vanish, and the assumed gauge-fixing is not correct. 
In fact, in the context of perturbation theory, there is no difficulty.
The operator $-\Delta$ (acting on functions that vanish at infinity) is strictly positive, and the $K^2$ term is nonnegative.\footnote{For a general choice of $S_\infty$,
the maximal volume Cauchy hypersurface has $K\not=0$, so it is not necessarily true that the $K^2$ term is perturbatively small.}   If we assume a strong energy
condition, then the $\h T_{tt}$ term is also positive, and this fact will be important in section \ref{class}.  But even if we do not assume a strong energy condition,
because of an explicit factor of $G$,  the $\h T_{tt}$ term is perturbatively small and does not affect the positivity of $\Xi$ in perturbation theory.

Now let us discuss the path integral $\int \d b \,\exp(\int_{S_0} b\Xi c^0)$ for the antighost field $b$.   
To do this integral, first recall that if $b$ and $c$ are odd variables and $A$ is a complex number, then $\int db \,\exp(bAc)=Ac = A \delta(c)$, since $c=\delta(c)$
for an odd variable.
Applying this principle on a mode-by-mode basis, we get
\be\label{yfred}\int Db \,\exp\left(\int_{S_0}\d^d x \sqrt{h} \,b\Xi c^0\right) =\det(\Xi) \delta(c^0|_{S_0}). \ee
The delta function of $c^0|_{S_0}$ has a simple meaning.  Since we have fixed $S_0$ to be the Cauchy hypersurface with maximal $V_R$, the remaining
gauge transformations that still have to be fixed are those that leave $S_0$ fixed (not necessarily pointwise, but as a set).   The restriction on the ghost field $c^\mu$
so that it generates a diffeomorphism that leaves $S_0$ fixed is precisely $c^0|_{S_0}=0$.   

To define the path integral in perturbation theory, one still needs a gauge-fixing condition for the remaining diffeomorphism group $\G_{S_0}$.   There is an unbroken subgroup
$\G_{\past}$ of diffeomorphisms that are nontrivial only to the past of $S_0$ (and in particular leave $S_0$ fixed pointwise); there is an analogous subgroup $\G_\fut$
consisting of diffeomorphisms that are nontrivial only to the future of $S_0$.   $\G_{S_0}$ is an extension of $\G_\past\times \G_\fut$ by the group $\diff\,S_0$
of diffeomorphisms of $S_0$:
\be\label{tellmex}1\to \diff\,S_0\to \G_{S_0}\to \G_\past\times \G_\fut\to 1.\ee
One may use any fairly standard gauge condition to fix $\G_{S_0}$.  One detail is that since we already have fixed the diffeomorphisms
that do not leave $S_0$ fixed, we do not need to fix those gauge symmetries again, and therefore we need a slightly smaller set of antighost fields and gauge
conditions than usual.   A convenient choice is to restrict the antighosts $\bar c^\mu$ by $\bar c^0|_{S_0}=0$.    Then $c^0$ and $\bar c^0$ are restricted 
in the same way, which makes possible a more natural-looking gauge-fixed action.  The details will not be important, however.

In quantum field theory in general, 
there is a standard strategy to factorize a transition amplitude
 on a spacetime $X$ by ``cutting'' on a Cauchy hypersurface $S\subset X$, as in fig. \ref{BOX}.   The goal of the cutting is to  express a path integral  on $X$ in terms of
states in a Hilbert space $\HH$ that consists of functions of the fields $\phi_S$ on $S$.   
Schematically, let $\Phi_S$ be the space of all possible values of the fields $\phi_S$.
  And for a given choice of $\phi_{S}$, let $\Phi_\past$ be the set of all fields to the past of $S$ and $\Phi_\fut$
 the set of all fields to the future of $S$.   The integral over $\Phi_\past$, keeping fixed the fields $\phi_{S}$ in $S$, 
 determines a ``ket'' vector $|\Psi_\past(\phi_S)\ra\in\HH$.
Similarly, the integral over $\Phi_\fut$, keeping fixed $\phi_{S}$, determines a corresponding ``bra'' vector $\la\Psi_\fut(\phi_S)|$.      Finally, one integrates over $\Phi_{S}$
to compute the inner product $\la\Psi_\fut|\Psi_\past\ra$.   This inner product gives the full path integral $Z_X$ over $X$, since by the time one integrates
over $\phi_S$, one has integrated over all fields to the past or future of $S$ or on $S$:
\be\label{polgo} Z_X=\la \Psi_\fut|\Psi_\past\ra= \int_{\Phi_S}  D \phi_{S}\, \bar\Psi_{\fut}(\phi_S)\Psi_\past(\phi_S). \ee  

Let us discuss how to implement this strategy in the present context,
with the above-described gauge-fixing which ensures that $S_0=S$ is the maximal volume hypersurface.   
First of all, in the gauge-fixing, we have ensured that $\K=0$ on $S$, so we cannot also fix the variable that is conjugate to $\K$.  This variable is the volume
density $\sqrt h$.   However, we are free to specify the conformal class of the metric on $S$.   Let us write $h_0$ for this conformal class; specifying $h_0$ is the
same as specifying $h$ up to a Weyl transformation $h\to e^{2\varphi}h$. 
Thus $h_0$ defines a point in $\Conf$, the space of conformal structures.   
 So a function of $h_0$ is a function on $\Conf$, and we formally denote the space of
such functions as
$\HH_\Conf$.     If matter fields are present, we can also specify the values of the matter
fields on $S$, and we  write $\HH_\matt$ for the Hilbert space of functions of  the matter fields.  Finally, we also have to consider
the ghosts.      The fields $c^0$ and $\bar c^0$ vanish along $S$, because of the conditions
$c^0|_S=\bar c^0|_S$ that were described earlier.  However, we do have fields $ c^i$ and ${\bar c}^i$ on $S$.   The functions of those
fields make up a ghost Hilbert space $\HH_\gh$.
The combined Hilbert space is then  $\HH_0= \HH_\Conf\otimes \HH_\matt\otimes \HH_\gh$.     (In a general situation, the definition of the ghosts and matter
fields might depend on $h_0$, and then a more precise statement is that $\HH_0$ is the combined Hilbert space of functions of
$h_0$, the matter fields, and the ghosts.)

In computing $\Psi_\past$, we perform a path integral to the past of $S$ with a boundary condition along $S$ that specifies the conformal structure $h_0$ of
$S$, and also specifies that $S$ has $\K=0$.   This is the mixed Dirichlet-Neumann boundary condition that was mentioned in section \ref{wdw} (now specialized
to $\K=0$).   It is elliptic, so the path integral that computes $\Psi_\past$ will be well-defined in perturbation theory.   The same is true for the path integral
that computes $\Psi_\fut$.

The inner product on $\HH_0$ is not the obvious one that would come from an integral over the fields $h_0, c^i,{\bar c}^j$ and possible matter fields.  
Rather,  an extra factor $\det\Xi$
 comes from the integral over $b$ in eqn. (\ref{yfred}).   Thus the inner product is formally
\be\label{bonus} \la \Psi_1|\Psi_2\ra =\int D h_0\,D  c^i\,D{\bar c}^j \,\, \bar\Psi_1 (\det\,\Xi) \Psi_2. \ee 
In the absence of the ghosts, this formula would define a positive-definite inner product on $\HH_\Conf\otimes \HH_\matt$, since 
the operator $\Xi$ is strictly positive and its determinant is therefore also positive.   However, the inner product on $\HH_\gh$ is not positive-definite.

At this point, we have to remember the BRST symmetry.  The whole gauge-fixing construction is BRST-invariant and leads to the existence of a BRST charge
$Q$ that acts on $\HH_0$.   The physical Hilbert space $\HH_\phys$ is defined as the cohomology of $Q$ acting on $\HH_0$.   In the context of perturbation theory,
passing to the BRST cohomology eliminates $c^i$ and ${\bar c}^j$ and also eliminates ``pure gauge'' modes of $h_0$.   Here  pure gauge modes
are the modes that are induced by diffeomorphisms of $S$.   The positivity of the underlying inner product on $\HH_\Conf\otimes \HH_\matt$ leads to positivity
of the inner product on $\HH_\phys$.  In the context of perturbation theory, to verify this one really only needs to know that positivity holds in the limit $G\to 0$ in which
all fields, including the ghosts, are treated as free fields.  Perturbative corrections will then not spoil this positivity.

  In the BRST formalism, the momentum constraint equation is satisfied because the generator
of the momentum constraint is a BRST commutator, $\P^i(x)=\{Q,\Lambda^i(x)\}$ for some operator $\Lambda^i(x)$,   This implies that $\P^i(x)$ acts
trivially on the BRST cohomology $\HH_\phys$, since if $Q\Psi=0$ then $\P^i(x)\Psi=Q(\Lambda^i(x)\Psi)$ vanishes as an element of $\HH_\phys$.   We do not
have to consider the Hamiltonian constraint, because we have eliminated it by considering a canonically determined Cauchy hypersurface $S_0=S$, the one
that has maximal renormalized volume.

In terms of the decomposition (\ref{tellmex}) of the residual gauge symmetry, the gauge-fixing of $\G_\past$ is a step in computing $\Psi_\past$,
the gauge-fixing of $\G_\fut$ is a step in computing $\Psi_\fut$, and the gauge-fixing of $\mathrm{diff}\,S_0$ is involved in constructing the BRST operator
$Q$ whose cohomology ultimately defines $\HH_\phys$.  In the context of perturbation theory, instead of relying on the BRST machinery, one could deal with the $\diff\,S_0$ symmetry by imposing gauge conditions that explicitly remove
the longitudinal modes of the metric $h$ of $S$.   
This would be analogous to axial gauge in gauge theory, and is one way to make manifest the positivity of the inner product on
$\HH_\phys$.   

In short, and modulo some  subtleties that are discussed later, 
we have arrived at a more precise version of the picture that was suggested heuristically in section \ref{constraint} based on facts about the classical
phase space: in  constructing a Hilbert space for AAdS gravity, at least in the context of perturbation theory, one can forget the troublesome Hamiltonian constraint
if one considers the quantum wavefunction to depend only on the conformal class $h_0$ of the metric, and not on the volume form.   
We also now know that to proceed in this way, one must include  a non-classical
factor $\det\,\Xi$ in the definition of the inner product.  

In AAdS gravity, this analysis enables us, at least in perturbation theory, to get a formula like that of eqn. (\ref{formula}) or fig. \ref{BOX}(a) in 
which a transition amplitude is factored
in terms of a sum over intermediate states on a Cauchy hypersurface.   The intermediate states are simply labeled by fields on the maximal volume Cauchy hypersurface
$S$.

In a similar fashion, one can
 get a formula like that of eqn. (\ref{formula2}) or fig. \ref{BOX}(b) in which an amplitude is written as a sum over states on a Cauchy hypersurface
$S_\infty$ in $X_\infty$, and also on another Cauchy hypersurface $S'_\infty$ to the future of $S_\infty$.   The bulk Hilbert spaces are defined on the
maximal volume Cauchy hypersurfaces $S$ and $S'$ with respective boundaries in $S_\infty$ and $S'_\infty$.   To extend the previous analysis to this case, we
 just need to know that if on the boundary $S'_\infty$
is everywhere to the future of $S_\infty$, then likewise in bulk $S'$ is everywhere to the future of $S$.   A simple argument for this is given in Appendix A of
\cite{Jacobson}.\footnote{Another proof can be deduced from  positivity properties of the operator $\Xi$.   Although there is no solution of $\Xi\epsilon=0$ that vanishes
at infinity, if one specifies a real-valued function $f$ on $S_\infty$, then there is a unique solution of $\Xi\epsilon=0$ on $S$ with $\epsilon\to f$ at infinity.
Moreover, if $f$ is positive, representing a first order displacement of $S_\infty$ into the future, then $\epsilon$ is also positive, representing a first order displacement  
of $S$ into the future. As $S_\infty$ moves into the future, at a rate determined by $f$, $S$ moves into the future, at a rate determined by $\epsilon$.}  Given
this, perturbative gauge-fixing such that two predetermined bulk hypersurfaces $S_0$ and $S_0'$ (with $S_0'$ to the future of $S$) both satisfy $\K=0$ (ensuring $S_0=S$, $S_0'=S'$) will lead to the desired factorization 
formula.   

Another generalization is as follows.\footnote{\label{evasion} This generalization, for suitable $\lambda$,  might enable one to circumvent the obstruction described in section \ref{compactification}.} Instead of gauge-fixing to require that $\K=0$ along $S_0$, we could pick an arbitrary real number $\lambda$ and gauge fix to require $\K+\lambda=0$ along $S_0$.
This is also a valid gauge condition, in the context of perturbation theory.  The analysis goes through much as before.
 Instead of being orthogonal to the boundary, as is the case if $\lambda=0$, $S_0$ will
now meet the boundary at a $\lambda$-dependent angle.   Since this introduces an asymmetry between future and past,
 it is most natural to now  view $\Psi_\past$ and $\Psi_\fut$ as vectors in dual,
$\lambda$-dependent spaces
$H_\lambda$ and $H_{-\lambda}$.   These spaces are not Hilbert spaces in a natural way, but there is a natural sesquilinear pairing
 $\la~,~\ra:\H_\lambda\times H_{-\lambda}\to \C$, and the path integral can be expressed in terms of this pairing, $Z_X=\la\Psi_\fut|\Psi_\past\ra$.
  From a classical point of view, as
 $\lambda$ varies from $-\infty$ to $\infty$, $S$ sweeps through the whole bulk domain of dependence $\Omega$ of $S_\infty$, from its past boundary to its future
 boundary.  It is not clear what is a useful quantum counterpart of this statement.

Now we will describe some subtleties concerning the definition of $\det\,\Xi$.  
To begin with, we discuss the dependence on $K$ and $h_0$.  
First consider the limit $G\to 0$.   We assume that the perturbation expansion is based on an expansion around some classical solution that is determined
by asymptotic conditions.    In this solution, $K$ is a $c$-number.    Moreover, the classical solution determines an actual metric on $S$, not just a conformal
class of metrics, so in the starting point of perturbation theory, there is a distinguished representative of the conformal class of metrics and we will write $h_0$ for this
representative.  Having a distinguished representative is important because the operator $\Delta$ is not conformally invariant.   In the classical limit, with $K$
and $h_0$ being given by the classical solution, the operator
 $\Xi= -\Delta+K^{ij} K_{ij}$ is a standard sort of second order differential operator, and its determinant $\det\,\Xi$ is 
 a fairly conventional functional determinant.   
 
 This determinant  arose in our derivation as the partition function of a theory
with a pair of fermi fields $b$ and $c=c^0$ on $S$ with action 
\be\label{tango} I_{bc}=-\frac{1}{2}\int_S \d^d x\sqrt h\,b(-\Delta+K^{ij} K_{ij}+8\pi G \h T_{tt}) c. \ee  
We think of $I_{bc}$ as the action of an auxiliary quantum field theory.
 Of course, in this limit, $\det\,\Xi$ is a highly nontrivial function of $g$ and   $K$.
But as soon as we turn on $G$-dependent corrections, $\det\,\Xi$ becomes something more interesting.   To explain this as simply as possible, consider
a model without matter fields, and suppose that
$S_\infty$ is such that the maximal volume hypersurface $S$, classically, has $K_{ij}=0$.   Then at $G=0$, $\Xi$ reduces to $-\Delta$.    But as soon
as we turn on perturbative corrections in $G$, the picture changes.   According to eqn. (\ref{momcon}), $K_{ij}$ is canonically conjugate to the metric tensor $h_{ij}$,
$K_{ij}=16\pi G\frac{1}{\sqrt h}\Pi_{ij}$ if $\K=0$.   $\Pi_{ij}$ acts as a derivative with respect to $h_{ij}$, and in the auxiliary quantum field theory with
action $I_{bc}$, this will give an insertion of the stress tensor $T_{ij}$.   Therefore, in first order, $K_{ij}K^{ij}$ becomes an insertion of
$G^2 T_{ij}T^{ij}$.   Thus the auxiliary quantum field theory undergoes a $T^2$ deformation, similar to the deformation considered in 
\cite{Verl,KLM, Taylor,HKST,Shyam,ALS}.  In a more general case, if $K$ is nonzero in the classical limit, we would interpret $K_{ij}$ as the sum of 
$-16\pi \i G\frac{\delta}{\delta h^{ij}}$ plus a classical contribution.
Inclusion of the matter fields in $\Xi$ gives a further deformation, as in \cite{AKW}, and there are also further corrections, as described shortly.

As usual, it is possible in principle to express the partition function in the deformed theory, to any finite order in perturbation theory, in terms of ordinary
correlation functions in the undeformed theory.  In the present case, this would be done by expanding the determinant in terms of the propagator of the operator
$-\Delta$ and insertions of the stress tensor.  However, because the perturbation is irrelevant in the renormalization group sense, as one goes to higher
and higher orders, one will encounter integrals that potentially have a very high degree of divergence and which require careful treatment.  Beyond perturbation theory, a definition
of the deformed theory is unknown.   This assertion is one aspect of the fact that the construction that we have given of a Hilbert space for AAdS gravity is,
in its present form, only valid in perturbation theory.

An important point here is that since we are specifying $h_0$ along $S$, the conjugate variable $K$ is not continuous along $S$ except in the classical limit, and will fluctuate independently
in the past and future of $S$.   The formula $K_{ij}=16\pi G\frac{1}{\sqrt h}\Pi_{ij}$ holds both to the past and the future of $S$; to the past of $S$, we interpret $\Pi_{ij}$ as a differential
operator that acts on the ket $|\Psi_2\ra$ in the inner product $\la\Psi_1|\Psi_2\ra$ that we are trying to calculate, while to the future of $S$, we interpret $\Pi_{ij}$ as a differential
operator that acts on the bra $\la\Psi_1|$.     This raises the question of how to interpret $K_{ij}$ when it appears in the operator $\Xi$ and seemingly must be evaluated precisely on
$S$.   The same question will arise in section \ref{kg} in the context of a Klein-Gordon particle, and there, since exact formulas are available, we can confirm that the obvious guess
is correct:   $K_{ij}$ along $S$ should be interpreted as the average of the values to the past and future of $S$.   Presumably something similar is true for gravity, though it would be harder 
to give a really convincing argument in the case of gravity.

Yet another question concerns the dependence of $\det\Xi$ on the conformal factor that appears in the metric $h=e^{2\varphi}h_0$.   Since $\det\,\Xi$ is not
conformally invariant, this dependence is nontrivial.  
As explained earlier, in the classical limit, we take for $h_0$ the actual metric determined by an underlying classical solution.   Then the combined data consisting of $h_0$
and the classical values of $K$ and the matter fields satisfy the Einstein equations and in particular satisfy the Hamiltonian constraint equation.  Quantum mechanically,
everything fluctuates, including the conformal factor $\varphi$ of the metric. The fluctuation in $\varphi$ is discontinuous across $S$, since we are fixing the conjugate variable
$\K$ to vanish along $S$.  However, if it is correct to assume that the fluctuations satisfy the Hamiltonian constraint equation,
then (on both sides of $S$)
 that equation determines the fluctuations in $\varphi$ in terms of the fluctuations in the conformal class of $h_0$ and $K$. Differently put, the Hamiltonian constraint, if valid, determines
a unique representative on each Weyl orbit.   Explaining this point is one of the main goals of section \ref{class}.   
Roughly speaking, we expect that on each side of $S$,  the Hamiltonian constraint equation
 remains valid and determines  $\varphi$  in terms of  $h_0$ and $K$.  Since we understand $h_0$ and $K$ as operators that act on the bra and ket wavefunctions,
  this makes it possible to interpret  $\varphi$ as such an operator  (giving a further correction to the $T^2$ deformation that was described earlier).   Why does the Hamiltonian constraint
  remain valid when the fields fluctuate?    If it is possible to put the path integral in canonical
 form near $S$, then the manipulation described in eqn. (\ref{zolto}) shows that the Hamiltonian constraint equation can be imposed near $S$.   But  even if we do not assume that
 this manipulation is valid, the vanishing of the Hamiltonian constraint operator $\H(\vec x)$ is the classical equation of motion for the metric component $g_{tt}$ that is ``normal" to $S$.
 So a multiple of $\H(\vec x)$ appearing in the functional integral -- for instance in $\det\,\Xi$ -- can be eliminated by redefining $g_{tt}$, and hence the Hamiltonian constraint can be used to
 eliminate $\varphi$ in favor of $h_0$ and $K$, and thus to replace $\varphi$ with a differential operator acting on the wavefunction.

The approach to constructing a canonical formalism that we have described is conceptually simple, as it is based on a simple gauge-fixing, 
but it has led to a variety of thorny technical questions, mostly concerning the understanding of the operator $\det\,\Xi$.  The best that we can say is that hopefully maintaining the BRST invariance of the construction determines unique answers to all these questions.

\subsection{Analogy With A  Klein-Gordon Particle}\label{kg}

Long ago, it was noted that the Hamiltonian constraint operator of gravity is formally a second order differential operator, somewhat like a Klein-Gordon operator
\cite{DeWitt}.   This motivated the suggestion that the inner product on solutions of the Wheeler-DeWitt equation might be analogous to a Klein-Gordon pairing.

For wavefunctions $\Phi_1,\Phi_2$ that satisfy the Klein-Gordon equation $(-g^{\mu\nu}D_\mu D_\nu +m^2)\Phi=0$ in a Lorentz signature spacetime $M$, 
the Klein-Gordon pairing is defined by
\be\label{pairing}(\Phi_1,\Phi_2)=\frac{\i}{2}\int_U \d \Sigma^\mu \,\bar\Phi_1\overset{\leftrightarrow}{\partial_\mu}\Phi_2 ,\ee
where $U$ is any Cauchy hypersurface in $M$ and $\bar\Phi_1\overset{\leftrightarrow}{\partial_\mu} \Phi_2=\bar\Phi_1\partial_\mu \Phi_2-\partial_\mu\bar\Phi_1\,
\Phi_2$.

One obvious problem with the analogy between gravity and Klein-Gordon theory is that the Klein-Gordon pairing is indefinite, while the Hilbert space inner product for gravity is supposed to be positive-definite.
Another obvious point is that the Hamiltonian constraint equation, which says that $\H(\vec x)=0$ for each point $\vec x$ in a Cauchy hypersurface $S$,
 is more similar to an infinite family of Klein-Gordon operators than to a single Klein-Gordon operator.
For the Klein-Gordon particle, the Klein-Gordon pairing is defined on a codimension 1 hypersurface $U$, so an analog of the Klein-Gordon pairing for gravity
should be defined on a submanifold of infinite codimension, with one constraint for each point in $S$.

That is essentially what we have done in defining the inner product (\ref{bonus}).   In gravity, it is often assumed that the wavefunction should be a function $\Psi(h)$
of the metric $h$ of an initial value surface $S$.  Thus such a wavefunction is a function on $\Met$, the space of metrics on $S$.
As explained in section \ref{wdw}, a drawback of such an approach is that the path integral that 
would formally compute a wavefunction $\Psi(h)$ (from given initial conditions and sources) is actually ill-defined, even in perturbation theory, since the requisite
boundary condition is not elliptic.   One may instead consider a wavefunction $\Psi(h_0,\K)$ that depends on a conformal structure $h_0$ on $S$ along
with a scalar function $\K$ on $S$ (interpreted classically as the trace of the second fundamental form of $S$ in a spacetime $X$).   The path integrals that
 compute wavefunctions $\Psi(h_0,\K)$ are well-defined in perturbation theory.
 
 The wavefunctions $\Psi_1$, $\Psi_2$ in the inner product that was defined in
 eqn. (\ref{bonus}) could be naturally defined as functions of $\K$ and $h_0$, but in the definition of the inner product, they are not integrated over $\K$ and $h_0$,
 but only over $h_0$, at $\K=0$.   This is analogous to the restriction from $M$ to $U$ in the Klein-Gordon pairing (\ref{pairing}): in the gravity case, as expected,  one
 places a condition at each point in $S$, namely $\K=0$.

Another detail  is that the symmetry under diffeomorphisms of $S$ is taken into account in eqn. (\ref{bonus}) not by asking for  $\Psi_1$ and $\Psi_2$ to be invariant under
the group 
$\Diff$ of diffeomorphisms of $S$,
but via the ghosts and the BRST formalism.     The difference is mainly important technically.   A Hilbert space of square-integrable 
functions or half-densities on an infinite-dimensional space  such as $\Met$ is a vague notion unless one can describe exactly what class of functions one is interested in.
In the BRST framework, the appropriate description is straightforward, at least in perturbation theory.   The BRST framework is not necessarily the only way to make perturbation
theory explicit -- for example, one could try to fix the pure gauge modes in $\Met$ by a sort of axial gauge -- but certainly the BRST machinery provides 
a simple framework for perturbation theory.

The last and crucial point about eqn. (\ref{bonus}) that requires some  elucidation is the factor $\det\,\Xi$.   In fact, we will now explain that this factor is quite analogous
to the factor $\overset{\leftrightarrow}{\partial_\mu}$ in the Klein-Gordon pairing.  In doing so, for brevity, we will take $M$ to be Minkowski space with metric
$\d s^2=\eta_{\mu\nu}\d X^\mu \d X^\nu=-\d T^2 +\d\vec X^2$, and we will take  $S$ to be the hypersurface $T=T_*$, for some $T_*$.   Generalizations are
straightforward.

The action for a Klein-Gordon particle in this spacetime can be described by a generally covariant theory on a one-dimensional worldline $\lambda$.   The metric
of the worldline is taken to be $g(t) \d t^2$, $g(t)\geq 0$, and we define $e(t)$ as the positive square root of $g(t)$.   The Klein-Gordon particle can then be
described by the action
\be\label{tellme} I=\frac{1}{2}\int_\lambda \d t \left(- e^{-1} \eta_{\mu\nu}\frac{\d X^\mu }{\d t}\frac{\d X^\nu}{\d t}-e m^2 \right),\ee
which is invariant under reparametrizations of $\lambda$.   The Hamiltonian constraint is the Euler-Lagrange equation for the field $e$; in other words, it is
$H=0$ with $H=-\frac{\delta{I}}{\delta e}=\frac{1}{2}\left(\frac{1}{e^2}\eta_{\mu\nu}\frac{\d X^\mu}{\d t}\frac{\d X^\nu}{\d t}+m^2\right)$.   Since the momentum conjugate to
$X^\mu$ is $\Pi_\mu=\frac{1}{e}\eta_{\mu\nu}\frac{\d X^\nu}{\d t}$, and upon quantization, $\Pi_\mu=-\i \frac{\partial}{\partial X^\mu}$, we have
\be\label{ohmo}H= \frac{1}{2}\left(\eta^{\mu\nu}\Pi_\mu\Pi_\nu+m^2\right) =\frac{1}{2}\left(-\eta^{\mu\nu}\frac{\partial}{\partial X^\mu}\frac{\partial}{\partial X^\nu}+m^2\right).
\ee

We now want to define states by conditions to the past and future of $S$, and define an inner product between them by some sort of integral on $S$.   In the spirit of
eqn. (\ref{formula}) or fig. \ref{BOX}(a), it would be natural to define initial and final states by conditions at $T=-\infty$ and $T=+\infty$.   However, a much shorter derivation
is possible if one is willing to define the states by means of sources at finite points to the past and future of $S$.   So we introduce points $X_0$ and $X_1$ respectively
to the past and future of $S$, at which states will be created and annihilated.  We can assume that $X_0$ has coordinates $X^\mu_0=(T_0,\vec X_0)$, $T_0<T_*$, and similarly $X_1$ has 
coordinates  $X^\mu_1=(T_1,\vec X_1)$, $T_1>T_*$.

Now we want to perform a path integral for the case that $\lambda$ is an interval, with boundary conditions such  that one end of the interval maps to $X_0$ and the other to $X_1$.
After evaluating this path integral, we will  explore how it can be factored in terms of states passing through the hypersurface $S$.

Because of reparametrization invariance, there is no loss of generality in assuming that $\lambda$ is the unit interval $0\leq t\leq 1$ with the endpoint $t=0$ mapped to
$X=X_0$ and the endpoint $t=1$ mapped to $X=X_1$.   The technique to do the path integral is well-known.   First of all, the length of the interval $\tau=\int_0^1\d t \,e(t)$
can be any positive number.   One can fix the reparametrization invariance of the interval by setting $e=\tau$.   The ghosts that are involved in this
gauge-fixing decouple.   For fixed $\tau$, the path integral over $X^\mu$ is just an ordinary quantum mechanical path integral on an interval of length $\tau$,
with the Hamiltonian $H$.   So the value of the path integral is $\la X_1|e^{-\i H\tau}|X_0\ra$.   
To evaluate the path integral, one has to integrate this matrix element over the remaining variable $\tau$ that is not determined by the gauge-fixing.
This integral is only conditionally convergent.  To define it precisely, one can include 
 a convergence factor $\exp(-\epsilon\tau)$ where $\epsilon$ is taken to 0 at the end of the calculation.  The output of the path integral is then
\be\label{output} G(X_1;X_0)=\int_0^\infty \d \tau \la X_1|e^{-\i H\tau - \epsilon \tau}|X_0\ra=\left\la X_1\left|\frac{-\i}{H-\i\epsilon}\right|X_0\right\ra. \ee
This obeys
\be\label{utput} H G(X_1;X_0)=-\i \delta^D(X_1-X_0), \ee
where one can consider $H$ to act either on $X_1$ or on $X_0$.   

Assuming that $m$ is large enough that $m(T_1-T_*),\, m(T_*-T_0)>>1$, $G(X_1,X_0)$ can also be computed in a perturbative expansion in which the starting point is
a solution of the classical equations of motion of this theory with the boundary conditions that $X=X_0$ at one endpoint and  $X=X_1$ at the other.    
There is a unique solution,\footnote{The  proper time elapsed in this solution 
is real if $X_1$ and $X_0$ are timelike separated, and imaginary if they are spacelike separated.  For an interesting analysis of 
the implications of this in the context of what in section \ref{wdw} was called the revised Wheeler-DeWitt formalism, see \cite{MarolfOther}.} 
namely a straight line 
trajectory from $X_0$ to $X_1$.    Such a trajectory, of course, intersects the hypersurface $U$ defined by $T=T_*$ in precisely one point.   Expanding around this
orbit, we learn that to all orders in an expansion in $1/m$, we can assume that a trajectory  intersects $U$ in a unique point.  

This means that, from the standpoint of perturbation theory in $1/m$, we can partially gauge fix the theory by requiring that some specified point on the interval $\lambda$
is mapped to $U$.   This step is analogous to the main step in section \ref{gaugefixing}, where we made a partial gauge-fixing to specify that a pre-chosen 
 hypersurface $S_0$ is the Cauchy hypersurface with $\K=0$.  
 
 To implement this idea in the present context, we can take $\lambda$ to be the interval $-1\leq t\leq 1$, with boundary conditions $X(-1)=X_0$, $X(1)=X_1$,
 and  a partial gauge-fixing condition $T(0)=T_*$.   To impose this condition, we  use the BRST formalism.   The BRST transformation of
 the field $X^\mu(t)$ is $\bigdelta X^\mu(t)= c\frac{\d X^\mu(t)}{\d t}$, where $c$ is the ghost field associated to an infinitesimal reparametrization of the worldline $K$.
 To implement the partial gauge-fixing, we introduce a BRST multiplet consisting of an antighost variable $b$ and a bosonic variable $\phi$ with BRST transformations
 \be\label{brstran} \largedelta b=\phi,~~~\largedelta\phi=0. \ee ($b$ and $\phi$ are defined only at $t=0$, so they are variables, not fields.) 
 The gauge-fixing action is 
 \be\label{gfix}\largedelta\left( b (T(0)-T_*)\right)= \phi (T(0)-T_*) -bc \frac{\d T(0)}{\d t}. \ee
 The integral over these variables is\footnote{To  properly justify the numerical factor $1/2\pi\i$ that we assume here in the measure would require a more
 precise derivation, possibly with a discretization of the path integral.}
 \be\label{nfix}\int \frac{\d b\,\d \phi}{2\pi\i}\exp\left(\i \phi(T(0)-T^*) -\i bc(0) \frac{\d T(0)}{\d t}\right)=-\delta(T(0)-T_*)\frac{\d T(0)}{\d t} \delta(c(0)) . \ee
 The delta function $\delta(c(0))$ means that $c(t)$ effectively splits up as two different fields, one of which is supported for $t<0$ and is associated to
 reparametrizations of the interval $-1\leq t\leq 0$, and one of which is supported for $t>0$ and is associated to reparametrizations of the interval $0\leq t\leq 1$.
 
 For a fixed value of  $\vec X(0)$, the path integral for $t\leq 0$ gives $G(T_*,\vec X(0);X_0)$ and the path integral over $t\geq 0$ gives $G(X_1; T_*,\vec X(0))$.
Integrating (\ref{nfix}) over $\vec X(0)$, we get the full path integral, which is supposed to equal $G(X_1;X_0)$, since we have merely analyzed  the
same path integral that led to eqn. (\ref{output}) with a different parametrization and gauge-fixing.   So we expect
\be\label{rfix} G(X_1;X_0)=-\int \d^{D-1}\vec X(0) \, G(X_1;T_*,\vec X(0)) \frac{\d T(0)}{\d t} G(T_*,\vec X(0); X_0). \ee
Here $\frac{\d T(0)}{\d t}$ can act as $-\i \partial_{T_*}$ on $G(T_*,\vec X(0); X_0)$, or as $+\i \partial_{T_*}$ on $G(X_1;T_*,\vec X(0))$.   The reason for the
relative minus sign is that the normal vector $\partial_t$ at $t=0$ is outward directed for the interval $-1\leq t\leq 0$ and inward directed for the interval $0\leq t\leq 1$.
The derivation that we are giving here is not precise enough to directly show whether $\frac{\d T(0)}{\d t}$ should be taken to act to the right or the left,
but the symmetry of the construction under exchange of the future and past shows that we presumably should take a symmetric combination of the two choices.
Thus we interpret the formula to be
\be\label{nfox} G(X_1;X_0) =-\frac{\i}{2} \int_U \d^{D-1} \vec X(0) \,G(X_1;T_*, \vec X(0)) \overset{\leftrightarrow}{\frac{\partial }{\partial T_*}} G(T_*,\vec X(0); X_0). \ee
In this formula, we see the Klein-Gordon inner product on the hypersurface $U$.  The formula says that a transition amplitude between states created
to the past and future of the hypersurface can be evaluated in terms of a sum over states on $S$, using the Klein-Gordon inner product.

To verify that this formula is in fact correct, let $\Theta(T_*-T(0))$ be the function that is $1$ for $T_*-T(0)>0$  and otherwise 0.
By using $\partial_{T(0)} \Theta(T_*-T(0))=-\delta(T_*-T(0))$, we can replace the integral over $U$ in eqn. (\ref{nfox}) with an integral over all of $M$:
\begin{align}\label{lfox}-&\frac{\i}{2} \int_U \d^{D-1} \vec X(0) \,G(X_1;T_*, \vec X(0)) \overset{\leftrightarrow}{\frac{\partial }{\partial T_*}} G(\vec X(0),T_*; X_0) \\
=&\frac{\i}{2} \int_M \d^{D-1} \vec X(0)\,\d T(0)\left( \frac{\partial}{\partial T(0)} \Theta(T_*-T(0)) \right)G(X_1;T(0), \vec X(0)) 
\overset{\leftrightarrow}{\frac{\partial }{\partial T(0)}} G(\vec X(0),T(0); X_0).\notag
\end{align}
Now we integrate by parts with respect to $T(0)$ and observe that for any functions $A,B$
\be\label{mogo} \partial_{T_0}( A\overset {\leftrightarrow} {\partial_{T_0}} B)=2\left( A(HB) -(HA)B \right)
+ \sum_{i=1}^{D-1} \partial_{X_i} (A\overset{\leftrightarrow}{\partial}_{X_i}B).\ee
When we use this in eqn. (\ref{lfox}), the terms $\partial_{X_i}(\cdots)$ can be dropped because we are integrating over $\vec X$ and nothing else depends on 
$\vec X$.   So the formula (\ref{lfox}) becomes
\be\label{rfox} G(X_1;X_0)={\i} \int_M \d^D X(0) \Theta(T_*-T(0)) \bigl(G(X_1;X(0)) H G(X(0); X_0) -(H  G(X_1;X(0)) ) G(X(0);X_0)\bigr), \ee
where $H$ acts on $X(0)$.
Finally, from (\ref{utput}), we have $H G(X(0);X_0)=-\i \delta^D(X(0)-X_0)$ and $HG(X_1;X(0)) =-\i \delta^D(X_1-X(0))$.  Of these two delta functions,
only the first is in the support of the function $\Theta(T_*-T(0)) $, and upon doing the integral, we confirm that eqn. (\ref{rfox}) is valid.

One surprise here is that although  the derivation of eqn. (\ref{nfox}) suggested that this formula is valid only in perturbation theory in $1/m$, the formula 
actually turned out to be exact.   It is not clear to what extent there is a general lesson here.

The derivation shows that the factor $\overset{\leftrightarrow}{\frac{\partial}{\partial X^\mu}}$ that makes the Klein-Gordon inner product indefinite can be interpreted
as coming from a ghost determinant.   For gravity, the analogous ghost determinant is $\det\,\Xi$, and is positive in perturbation theory.

\section{The Classical Phase Space}\label{class}

In section \ref{orient}, we explained just enough about the relation of the classical phase space $\vP$ of AAdS gravity to a cotangent bundle $T^*(\Conf/\Diff)$ to motivate
the quantum treatment in section \ref{gaugefixing}.    Here we will give a more complete explanation and also a more general one, including matter fields.

In this discussion, $\Conf$ is the space of conformal structures on a Cauchy hypersurface $S$ in a spacetime $X$, and $\Diff$ is the group of diffeomorphisms of $S$.  If
$X$ is asymptotically Anti de Sitter, which is our main focus, 
 the conformal structure on $S$ is required to be asymptotic to a specified conformal structure on the boundary $S_\infty$, and diffeomorphisms of $S$
are required to be trivial at infinity.  However, some of the considerations can be adapted to a closed universe -- that is, to the case that $S$ is compact.

To establish an equivalence of $\vP$ to $T^*(\Conf/\Diff)$, or a generalization of this to include matter fields, one finds maps in both directions that are inverses
of each other.   The map from $\vP$ to $T^*(\Conf/\Diff)$ is made by finding a maximal volume hypersurface with specified asymptotic behavior, and the map
in the opposite direction is made by solving the Lichnerowicz equation to find a Weyl factor by means of which the Hamiltonian constraint equation is satisfied.
The two maps are inverses of each other, under appropriate conditions, and this establishes the isomorphism between $\vP$ and $T^*(\Conf/\Diff)$.
We begin by discussing the maximal hypersurfaces and then we consider the Lichnerowicz equation.

\subsection{Maximal Hypersurfaces}\label{extremal}

\subsubsection{Extremal  Hypersurfaces and Maximal Ones}

The first important fact is that in pure gravity with negative cosmological constant, and also in the presence of matter fields
that satisfy the strong energy condition, a hypersurface of extremal volume is automatically a local maximum of the volume.
To be more precise, we consider a Cauchy hypersurface $S\subset X$ that is asymptotic at infinity to some given Cauchy hypersurface  $S_\infty\subset X_\infty$,
and we assume that $S$ has extremal renormalized volume among all Cauchy  hypersurfaces that are asymptotic to $S_\infty$.  The claim is that, in a large class of theories, the renormalized
volume $V_R(S)$ is actually a local maximum among this class of hypersurfaces.

 As in section \ref{gaugefixing}, we can pick local coordinates $t, \vec x$ near $S$ so that $S$ is defined by $t=0$ and the metric near $S$ takes the form 
\be\label{meltform}\d s^2 =-\d t^2 +\sum_{i,j=1}^d g_{ij}(\vec x,t) \d x^i \d x^j. \ee
Consider a nearby hypersurface $S'$ defined by $t=\epsilon(\vec x)$ for some function $\epsilon$.   We require that $\epsilon(\vec x)$ vanishes at infinity so that $S$ and 
$S'$ are asymptotic to the same boundary hypersurface $S_\infty$.   

 If $S$ is an extremum of the renormalized volume,
then the renormalized volume of $S'$ coincides with that of $S$ in order $\epsilon$, and the $\epsilon^2$ term was identified in eqn. (\ref{winco}): 
\be\label{wincony}V_R(S')=V_R(S)-\int_{S}\d^dx \sqrt{h}
\left( \frac{1}{2}h^{ij}\partial_i \epsilon(\vec x) \partial_j \epsilon(\vec x) +\frac{1}{2}\epsilon(\vec x)^2 \left(8\pi G \h T_{tt}+K_{ij}K^{ij}\right)\right)+\O(\epsilon^3),\ee
where $\h T_{\mu\nu}=T_{\mu\nu}-\frac{1}{D-1} g_{\mu\nu} T^\alpha_\alpha$, with $T_{\mu\nu}$ the matter stress tensor 
(including a contribution from the cosmological constant).   We see that if $\h T_{tt}\geq 0$,
then the $\epsilon^2$ term in $V_R(S')$ is negative (the condition that $\epsilon\to 0$ at infinity ensures that the  $-(\nabla \epsilon)^2$ on the right hand
side of eqn. (\ref{wincony}) is strictly negative for any $\epsilon\not=0$).     This shows that in this AAdS context, assuming that $\h T_{tt}\geq 0$, an extremum of the renormalized volume is always a local
maximum.    

The condition that $\h T_{tt}\geq 0$ at each point and in each local Lorentz frame is called the strong energy condition.
Partly because of its role in the argument just sketched, the strong energy condition is important in relating the phase space of AAdS gravity to a 
cotangent bundle.\footnote{We will see that the same condition is also important in analyzing the Lichnerowicz equation.} 
   In what theories does it hold?  It holds for pure gravity with negative cosmological
constant, and it holds in any dimension for gravity coupled to $p$-form fields, $p\geq 1$, and to scalar fields with a non-positive potential.      
For example,  the strong energy condition holds
in all of  the usual 10 and 11 dimensional supergravity theories with the exception of the massive Type IIA supergravity theory, which was constructed in \cite{Romans}.   
 These facts are explained in section \ref{pform}.   The outstanding example of a theory that does {\it not} satisfy the strong energy condition is gravity with a positive
 cosmological constant, and more generally, any theory that contains scalar fields in which the scalar potential is not negative semi-definite.

In a model that satisfies the strong energy condition, 
the fact that any extremum of the renormalized volume  is a local maximum suggests that the extremum of the renormalized volume is unique:
viewing the renormalized volume as a function on the space of Cauchy hypersurfaces with specified asymptotics, between two local maxima one 
would 
 expect to find a saddle point, contradicting the fact that every extremum is a local maximum.
A proof of this uniqueness was given in Appendix A of \cite{Jacobson} by  use of the Raychaudhuri equation.\footnote{For $X={\mathrm{AdS}}_3$, another
proof of uniqueness is given in \cite{BoS}.   This proof is valid even if the conformal boundary $S_\infty$ of $S$
 is highly nonsmooth (which complicates the definition and analysis of $V_R$).  
That case is important for some applications (for example, see \cite{SK}), but for our purposes in the present article, we can assume that $S_\infty$ is smooth.}
For completeness, we will summarize the argument (the details are not needed for the rest of this article).   Let $S$ be an extremal 
Cauchy surface whose uniqueness we wish to prove, and let $S'$ be some other Cauchy surface with the same asymptotic behavior.  Given a point $p\in S$, let $\gamma_p$
be the geodesic through $p$ that is normal to $S$.   By global hyperbolicity, $\gamma_p$ intersects $S'$ at a unique point $p'$;
let $\gamma_{[p,p']}$ be the segment of $\gamma_p$ from $p$ to $p'$.  Then 
$\gamma_{[p,p']}$  may or may not be the causal path  from $p$ to $p'$ that has the greatest possible elapsed proper time.  
Let $S_0$ be the subset of $S$ consisting of points $p$ such that
$\gamma_{[p,p']}$ is proper time maximizing.  Define $\varphi_0:S_0\to S'$ by $\varphi_0(p)=p'$ if $p'=\gamma_p\cap S'$.   A standard argument (see for example \cite{Wald,Wittenrays})
using global hyperbolicity and compactness of spaces of causal paths shows that every point $p'\in S'$
can be reached from $S$ by a causal path that maximizes the elapsed proper time; moreover, this path is a geodesic orthogonal to $S$ at some point $p\in S$.  So the
map $\varphi:S_0\to S'$ is surjective.   Moreover, if $\h T_{tt}$ is everywhere strictly positive, Raychaudhuri's equation implies that 
$\varphi$ is everywhere volume-reducing.
Hence the volume of $S'$ is strictly less than the volume of $S_0$, and this in turn is no greater than the volume of $S$.   So $S$ has greater volume than any other
Cauchy hypersurface. So any extremal Cauchy hypersurface is a strict maximum of the renormalized volume, and is  therefore unique.

\subsubsection{Existence of a Local Maximum}\label{exi}

The next step is to discuss the {\it existence} of a local maximum of $V_R$, in the AAdS setting, for hypersurfaces with specified asymptotic behavior.   
Here what is known is actually incomplete. A detailed discussion  leads almost
inevitably to questions about cosmic censorship and the singularities of classical solutions of Einstein's equations.

For the case $X=\AdS_D$, for any dimension $D$, and any choice of $S_\infty$, a proof of existence of an extremal  Cauchy hypersurface $S\subset X$ with boundary $S_\infty$
 was given in \cite{BoS}.\footnote{In addition, using this result, existence and uniqueness of an extremal Cauchy hypersurface
 was shown in \cite{SK} for a spacetime that is locally (not just
 asymptotically) $\AdS_3$, in other words, for any classical solution of pure Einstein gravity in $D=3$ with $\Lambda<0$.
   This result holds for arbitrary topology of the initial value surface; in particular, the boundary may have any number of connected
 components.}   More recently \cite{CG}, existence of such a hypersurface was shown in any AAdS spacetime  under the hypothesis that the bulk domain of dependence of $S_\infty$ is compact.
 One goal of the following qualitative remarks is to explain the role of that assumption; the other goal is to explain that under the same assumption, one should expect a similar
 result for $\AdS_D$ compactifications, that is, for  spacetimes that are asymptotic to $\AdS_D\times W$ for some compact $W$.

Let us say that a Cauchy hypersurface\footnote{We are about to make an argument that involves limits. A sequence of spacelike hypersurfaces can develop
null portions in a limit.   So technically, in the following argument, it is best to define a Cauchy hypersurface to be
 a complete achronal, but not necessarily spacelike, hypersurface
on which initial data can be formulated; it may have null portions.  The null portions have zero volume so  a volume-maximizing hypersurface
will not  have null portions.}  in a spacetime $X$ that is asymptotic to $\AdS_D$ or $\AdS_D\times W$   is ``allowed'' if it is asymptotic to some chosen boundary Cauchy hypersurface $S_\infty\subset X_\infty$.
Any allowed Cauchy hypersurface is contained
in  the bulk domain of dependence\footnote{The bulk domain of dependence of $S_\infty$ is the domain of dependence of any allowed bulk hypersurface $S$;
alternatively, it is the set of points in $X$ that are not timelike separated from $S_\infty$.   Technically, in the following argument, it is 
convenient to include the points of $S_\infty$ in $S$ and in the bulk domain of dependence $\Omega$; this ensures that
$S$ is compact, and makes it possible for  $\Omega$  to be compact (as we wll see, this happens if $X$ is geodesically complete).}  
of $S_\infty$, which we will call $\Omega$. We assume that $\Omega$ is compact; this assumption will be discussed critically later.  Now let  $S_1,S_2,\cdots $ be a sequence of 
allowed Cauchy hypersurfaces.
The $S_i$  cannot go to infinity in spacetime, since they are all contained in the compact
set $\Omega$.  The condition that they all are everywhere spacelike or null means that they also cannot go to infinity in momentum space.  More specifically, if
a hypersurface $S$ is described locally by specifying a function $t=f(\vec x)$ where $\vec x$ and $t$ are local space and time coordinates, then the condition
for $S$ to be spacelike or null is $|\nabla f|\leq 1$, which is a sort of momentum space bound.   Because the $S_i$ are bounded in both position space and momentum space,
the sequence $S_i$ has a  (pointwise) convergent subsequence.\footnote{\label{wall}For  a fuller explanation of this type of argument about sequences of hypersurfaces, 
 see the proof of Theorem 10 in \cite{Wall}.  An important detail is that the renormalized volume is only upper semicontinuous on the space of Cauchy hypersurfaces that
 are asymptotic to $S_\infty$, meaning that in a limit, it can jump upward but cannot jump downward.  However, since we are trying to maximize the renormalized volume,
 upward jumps are not a problem.
 (To see why upward jumps in volume are possible, consider a sequence of spacelike 
 hypersurfaces that look locally like $t=\frac{1}{2}\epsilon \cos(x/\epsilon)$, where $x$ is
 a space coordinate and $\epsilon<<1$.  This family of hypersurfaces has a limit
 for $\epsilon\to 0$, namely the hypersurface $t=0$, and the volume jumps upward at $\epsilon=0$.)}  The renormalized volume is bounded above\footnote{Although $V_R$ is bounded above on the space of allowed Cauchy hypersurfaces, it is actually not
 bounded below.   If $S$ is asymptotic to $S_\infty$ but is not  orthogonal to $X_\infty$, then $V_R(S)=-\infty$.  That is because the unsubtracted volume $V(S)$ is
 always $+\infty$, so an infinite subtraction has been made to define $V_R(S)$.  A hypersurface that is not orthogonal to the boundary has a less divergent volume
 than an orthogonal one, so its renormalized volume is $-\infty$.   An allowed hypersurface that is orthogonal to the boundary has a finite renormalized volume,
 but the renormalized volume of such hypersurfaces can be arbitrarily negative, since a sequence $S_1,S_2,\cdots$ 
 of allowed hypersurfaces that are orthogonal to $X_\infty$
 might have a limit that is not orthogonal to $X_\infty$; in that case $\lim_{i\to\infty}V_R(S_i)=-\infty$.}
  as a function on the space of allowed Cauchy hypersurfaces, since a sequence $S_1,S_2,\cdots$ with $V_R(S_i)$ tending
 to $+\infty$ could not have a convergent subsequence.   Let $V_\max$ be the least upper bound on $V_R(S)$ among allowed Cauchy hypersurfaces $S$,
 and consider a sequence $S_1, S_2,\dots$ of allowed hypersurfaces with $\lim_{i\to\infty} V_R(S_i)= V_\max$.   The limit $S$ 
 of a convergent subsequence of the sequence
 $S_1,S_2,\cdots$ will have $V_R(S)=V_\max$ and will be a maximal volume  hypersurface.

A key assumption in this argument was that the bulk domain of dependence $\Omega$ is compact.
This is true if $X=\AdS_D$, but in a general   spacetime that is asymptotic to $\AdS_D$, 
 $\Omega$ may fail to be compact, 
 because singularities may form in the  evolution of $X$ from initial data on $S$.   For example, if a Schwarzschild black
hole forms to the past or future of $S$, the domain of dependence of $S$ may not be compact.   However, the presence of a Schwarzschild singularity
does not spoil the existence of a volume-maximizing hypersurface, for the following reason.  
A Schwarzschild singularity is a special case of a more general type of singularity known as a Kasner singularity.
 A Kasner singularity is a solution of Einstein's equations of the form\footnote{This is a solution of Einstein's equations with zero cosmological constant;
 however, the cosmological constant is not important near the singularity.}
\be\label{kasner} \d s^2= - \d t^2 +\sum_{j=1}^d t^{2p_j}  (\d x^j)^2,~~~\sum_{i=1}^d p_i=\sum_{i=1}^d p_i^2 =1.\ee
The volume form of a hypersurface $t=t_0$ vanishes as $t_0$  approaches the singularity at $t=0$, so a volume-maximizing hypersurface
is repelled from a Kasner singularity.     (Essentially this point is discussed in \cite{Wall} in the proof of Theorem 11.)  
Therefore, noncompactness of $\Omega$ due to formation of a Kasner singularity poses no problem for the existence of a volume-maximizing hypersurface.
A Schwarzschild singularity is the
special case of a Kasner singularity with one of the $p_i$ equal to $-(d-2)/d$ and the others equal to $2/d$, so it causes no difficulty.
Formation of a Kerr black hole  causes no difficulty
because the singularity of a Kerr black hole is timelike and would not be contained in the domain of dependence $\Omega$.  Belinski-Khalatnikov-Lifshitz (BKL) singularities are similar to Kasner singularities but, roughly, with repeated jumps in the exponents as $t\to 0^+$, so one would expect them to cause no difficulty.

It is conjectured that generic spacelike singularities in General Relativity are of BKL  type \cite{HenneauxBook}.
   Under this assumption,
 we can hope that a maximal volume hypersurface $S$ with specified asymptotic behavior always exists in 
any asymptotically AdS$_D$ spacetime, for any $D$.   In a theory in which  the  strong energy condition holds,  $S$ would be unique.

We should caution the reader, however, that compactifications to AdS$_D$ are different.  In a spacetime $X$ asymptotic not to AdS$_D$
but to $\AdS_D\times W$ for some compact manifold $W$, we will in section \ref{compactification} explain a simple argument showing in some cases
that a maximal volume hypersurface cannot exist.   This will actually motivate a conjecture that non-BKL singularities form generically in such compactifications.

We will now show that assuming the null energy condition,\footnote{The null energy condition states that at each point $q\in X$ and for each null vector $n$, 
the stress tensor $T$ satisfies $n^\alpha n^\beta T_{\alpha\beta}(x)\geq 0$.  This condition holds rather generally in physically sensible relativistic field theories
(in theories with scalar fields, it holds in Einstein frame).} if $X$ is free of singularity, or more precisely if it is geodesically complete, then $\Omega$ is compact.  (This argument is not needed in the rest
of the article.)  First we  show that if $X$ is
geodesically complete, then the future boundary of $\Omega$, which we will denote as $\partial_+\Omega$, is compact; similarly the past boundary $\partial_-\Omega$
is compact.    The reasoning involved is similar to that in the proof of Penrose's
singularity theorem; see for example \cite{Wald} or \cite{Wittenrays}.  
Any point $p\in\partial_+\Omega$ can be reached from $S_\infty$ by an orthogonal null geodesic
without a focal point.   The  future-going inward orthogonal null geodesics that originate on $S_\infty$ are initially converging, and 
assuming the null energy condition (which holds in reasonable classical field theories),
 Raychaudhuri's equation implies
that if $X$ is geodesically complete, they all reach focal points, beyond which they are not contained in $\partial_+\Omega$.
  The segment of any such geodesic that is contained in $\partial_+\Omega$ (including its initial point
on $S_\infty$) is therefore compact, and as $S_\infty$ is also compact, it follows that $\partial_+\Omega$ is compact.   Similarly $\partial_-\Omega$ is compact.   
Given this, to show that $\Omega$ is compact, we can for example use the fact that a globally hyperbolic manifold $X$ with
Cauchy hypersurface $S$ can be put in the form $S\times \R$ where the set $p\times \R$ is timelike for any $p\in S$, and $\R$ is parametrized by a variable $u$
that, for each $p\in S$, runs over the full range $-\infty<u<\infty$.   Compactness of $\partial_+\Omega$ and $\partial_-\Omega$ implies that the function
$u$ is bounded on $\partial_+\Omega$ and on $\partial_-\Omega$.   For $p\in S$, let $u_+(p)$ be the least upper bound of 
$u$ on $(p\times \R)\cap \Omega$, and similarly
let $u_-(p)$ be the greatest lower bound of $u$ on $(p\times \R)\cap \Omega$.   Then $\Omega$ consists of 
points $p\times u\in S\times \R$ with $u_-(p)\leq u \leq u_+(p)$,
and so is compact.

\subsection{The Phase Space and the Constraint Equations}\label{phase}

The existence and uniqueness of a maximal volume hypersurface $S$, discussed in section \ref{extremal}, is one ingredient
in relating the phase space of AAdS gravity to a cotangent bundle
$T^*(\Conf/\Diff)$.   The other ingredient, as developed in  \cite{Monc,KS,BoS,SK} for the case $D=3$, involves analyzing the Einstein constraint equations and in particular showing that the Hamiltonian constraint equation
can be viewed as a condition that fixes the Weyl factor in the metric  of a Cauchy hypersurface $S$.  Here, we will explain this
argument for the case of pure gravity with negative cosmological constant. Matter fields will be included in section \ref{matf}.

Suppose that $S$ is a Cauchy hypersurface in a spacetime $X$ of dimension $D=d+1$ that satisfies Einstein's equations with negative cosmological constant $\Lambda$.
The metric $h$ and second fundamental form $K$ of $S$ automatically satisfy the Einstein constraint equations:
\begin{align}\label{coneq} D_i K^{ij}-D^j K^i_i & = 0 \cr
                            R(h)& =K^{ij} K_{ij}-K^i_i K^j_j+2\Lambda, \end{align} where $R(h)$ is the scalar curvature of the metric $h$.
 These equations were introduced previously in section \ref{constraint}; the first is called the momentum constraint and the second is the Hamiltonian constraint.
  Any  pair $h,K$  satisfying these constraint equations on a manifold $S$ provides  initial data that determines 
 a spacetime $X$ that satisfies Einstein's equations and has $S$ as a Cauchy hypersurface.      
 
 Given suitable assumptions about singularities as discussed in section \ref{exi}, we expect that if $X$  is an AAdS solution of Einstein's equations, there is a unique
 volume-maximizing Cauchy hypersurface $S\subset X$  asymptotic to any given boundary Cauchy hypersurface $S_\infty$. 
 The metric $h$ and second fundamental form $K$ of $S$ satisfy the Einstein constraint equations (with $K^i_i=0$, since $S$ is volume-maximizing),        
 and the spacetime $X$ can be recovered from $S$ by solving Einstein's equations with initial data $h,K$.   Since $S$ is unique, two spacetimes
 obtained this way are equivalent if and only if they are equivalent via a diffeomorphism of $S$.    So in short, under the given
 assumption about singularities, the phase space $\vP$ of solutions of the  Einstein equations in the domain of dependence of a boundary Cauchy hypersurface
 $S_\infty$
 is the same as the space of solutions of the constraint equations (\ref{coneq}) with $K^i_i=0$ and $S$ asymptotic to $S_\infty$, up to diffeomorphism of $S$.  
Our goal here is to show  that this space  is   $T^*(\Conf/\Diff)$, implying that $\vP=T^*(\Conf/\Diff)$, under our assumptions.

As a first step,  observe that once we set $K^i_i=0$,  which reduces the momentum constraint to $D_i K^{ij}=0$, 
the momentum constraint becomes  Weyl-invariant.   To be precise, if we introduce a Weyl-rescaled metric
\be\label{olygo} \t h = \phi^\ell h,\ee
 with a positive function $\phi$, and similarly rescale the second fundamental form, setting
\be\label{polygo} \t K^{ij}=\phi^{-\ell(1+d/2)}K^{ij}, \ee
then the momentum constraint simply becomes
\be\label{zonteq} \t D_i \t K^{ij}=0, \ee
where $\t D_i$ is the covariant derivative computed with the new metric $\t h$.   
Though this assertion is easily verified, it may seem mysterious at first sight,
since the Weyl rescaling in question is certainly not a symmetry of General Relativity.  
In section \ref{symplectic}, we will give a more conceptual explanation of this Weyl invariance, but here we explain why it is useful.

The point is that if we are given a pair $h,K$ that satisfies the momentum constraint equation (and the AAdS boundary condition at infinity), then
there is a unique Weyl transform of this pair that satisfies the Hamiltonian constraint equation.    Thus for the purposes of describing
the phase space, we can simply replace the Einstein constraint equation with the operation of dividing by the group $\W$ of Weyl transformations.
Therefore,  the only constraint equation that we have to discuss explicitly is the momentum constraint; the Hamiltonian constraint can be replaced by
the group of Weyl transformations. This is a substantial
simplification, because the momentum constraint is linear in $K$, and is much easier to understand than the Hamiltonian constraint equation.  

Thus, the key fact is that given a pair $\t h,\t K$ that satisfies the momentum constraint equation, there is a unique positive function $\phi$ such that
the Weyl rescaled metric and second fundamental form $h=\phi^{-\ell} \t h$, $K^{ij}=\phi^{\ell(1+d/2)}\t K^{ij}$ satisfy the Hamiltonian 
constraint equation.  In sketching the proof,        
 we will follow the very useful explanation in section 3 of \cite{crusc}.  For convenience, we use the  notation of that paper.   See also, for example,
 \cite{OY,Ch,CI,I,BI}.

Setting   $\ell=-4/(d-2)$, the scalar curvatures $R(h)$ and $R(\t h)$ are related by
\be\label{zimbo} R(h)\phi^{\frac{4}{d-2}} =R(\t h)-\frac{4(d-1)}{(d-2)\phi}\Delta_{\t h}\phi, \ee
where $\Delta_{\t h}=\t h^{ij}\t D_i \t D_j$ is the Laplacian for the metric $\t h$.  An important preliminary point  is that in an AAdS spacetime, this equation can be used
to show that, by a Weyl transformation that is trivial at infinity, we can set $R(\t h)=2\Lambda$ everywhere (not just at infinity) \cite{ACF}.
Actually, in the following argument, it suffices for $R(\t h)$ to be negative-definite, and knowing that this suffices is important 
background for understanding what happens when
matter fields are included.  So we will retain $R(\t h)$ in the formulas and assume only that it is negative, and  approaches $2\Lambda$ at infinity.  Since we want also
$R(h)\to 2\Lambda$ at infinity, we can assume that the function $\phi$ approaches 1 at infinity.

We perhaps should stress at this point that the ability to make a Weyl rescaling to set $R(\t h)<0$ is special to an AAdS spacetime.  There are potential obstructions to this
in a closed universe, and the statement also has no equally simple analog for gravity with zero or positive cosmological constant.
That is one of the reasons that the conformal approach to the constraint equations, which we are describing here, is particularly powerful in an AAdS spacetime.   
Another reason is that the arguments of section \ref{exi} concerning maximal hypersurfaces in AAdS spacetimes do not have equally satisfactory analogs in other cases.

 Let us define $|\t K|^2_{\t h}=\t K^{ij}\t K^{i'j'}\t h_{ii'}\t h_{jj'}$. (Similar notation will be used later for other tensors.)  Making use of eqn. (\ref{zimbo}), we find that the
 Hamiltonian constraint equation   in (\ref{coneq}) becomes
 \be\label{poneq} \Delta_{\t h} \phi -\frac{(d-2)}{4(d-1)} R(\t h) \phi+\frac{(d-2)}{4(d-1)}|\t K|^2_{\t h} \phi^{{(2-3d)}/{(d-2)}} +\frac{\Lambda(d-2)}{2(d-1)}\phi^{{(d+2)}/{(d-2)}}=0.\ee
 In this form, the Hamiltonian constraint is called the Lichnerowicz equation.
It can be written
 \be\label{zoneq}\Delta_{\t h}\phi-F(\phi,x)=0,\ee with
 \be\label{woneq}  F(\phi,x)=\frac{(d-2)}{4(d-1)}R(\t h) \phi-\frac{(d-2)}{4(d-1)}|\t K|^2_{\t h} \phi^{(2-3d)/(d-2)} -\frac{\Lambda(d-2)}{2(d-1)}\phi^{(d+2)/(d-2)}.\ee
 Here $x$ denotes a point in $S$, and the explicit $x$-dependence of $F(\phi,x)$ comes from the $x$-dependence of $R(\t h)$ and $|\t K|^2_{\t h}$.
 
 To complete the description of the phase space, we want to show that, with $R(\t h)$ and $\Lambda$ both negative, 
 there is a unique positive function $\phi$ that satisfies the
 Lichnerowicz equation and approaches 1 at infinity.  The main tool is the following.  A positive function $\phi_-$ is called a subsolution if  the left hand side of eqn. (\ref{zoneq}) is nonnegative,
 \be\label{goneq} \Delta_{\t h}\phi-F(\phi_-,x)\geq 0,\ee
 and a positive function $\phi_+$ is called a supersolution if the right hand side is nonpositive,
 \be\label{honeq}\Delta_{\t h}\phi_+ -F(\phi_+,x)\leq 0.\ee
 If there is a subsolution $\phi_-$  and a supersolution  $\phi_+$ with $\phi_-\leq \phi_+$, then we will prove that there exists a solution $\phi$ of the Lichnerowicz
 equation with
 \be\label{bonc} \phi_-\leq \phi\leq \phi_+.\ee
 
 For small $\phi$, the dominant term in $F(\phi,x)$ is the $|\t K|^2_{\t h}$ term wherever $\t K\not=0$, and the $R(\t h)$ term wherever $\t K=0$.  Both
 of these contributions to $F(\phi,x)$ are negative, since we have assumed $R(\t h)<0$.   So an example of a subsolution is any sufficiently small constant $C$.
 For large $\phi$, the dominant term in $F(\phi,x)$ is the term proportional to $-\Lambda$. We have assumed $\Lambda<0$, so this term is positive.
 Hence an example of a supersolution is any sufficiently large constant $D$.   On a compact manifold $S$, these choices of subsolution and supersolution 
 are satisfactory.   However, in the AAdS context, we want a solution of the Lichnerowicz equation such that $\phi\to 1$ at infinity.   Eqn. (\ref{bonc}) 
 will guarantee this if $\phi_-$ and $\phi_+$ both approach 1 at infinity.  So we want a subsolution and a supersolution with that property.
 These can be found as follows.  Write the AAdS metric in the familiar form
 \be\label{pko}\d s^2=\frac{1}{\rho^2} \left(\d  \rho^2 +\sum_{a,b=1}^d g_{ab}(x,\rho)\d x^a\d x^b\right),\ee
 where the conformal boundary $S_\infty$ is at $\rho=0$.   For a suitably small constant $\epsilon$, take $\phi_-$ to equal 1 at $\rho=0$ and $C$ for $\rho\geq \epsilon$,
 with a smooth monotonic interpolation in between, and similarly take $\phi_+$ to interpolate from 1 at $\rho=0$ to $D$ for $\rho\geq \epsilon$.   With a suitable choice
 of the interpolations, this gives a subsolution and a supersolution with $\phi_-\leq \phi_+$ everywhere and $\phi_-,\phi_+\to 1$ at $\rho=0$.   See section 5 of
 \cite{Sak} for a more detailed explanation of this point.
 
 The proof of existence of a solution of the Lichnerowicz equation, given $\phi_-$ and $\phi_+$, is simpler  on a compact manifold, so we begin
 with that case.   
 For a sufficiently large constant  $c$, the function $F_c(\phi,x)=F(\phi,x)-c\phi$ is (for any $x$) monotone decreasing
for $\phi$ in the interval\footnote{This statement is also true
 in the AAdS case, because natural AAdS boundary conditions ensure that all coefficients in $F(\phi,x)$ are bounded at infinity; indeed, $R(\t h)$ is asymptotically
 constant, and $|\t K|^2_{\t h}\to 0$ at infinity.} $[C,D]$.     The operator $\Delta_{\t h}-c$ is negative-definite, so for any function $f$,
 the equation $(\Delta_{\t h}-c)\phi=f$ has a unique solution for $\phi$.   So we can inductively define a sequence of functions $\phi_0,\phi_1,\cdots$
 with $\phi_0=\phi_+$ and for $n\geq 1$, 
 \be\label{thestep}(\Delta_{\t h}-c) \phi_{n} =F_c(\phi_{n-1},x).\ee
Suppose that  for all $n\geq 0$, 
 \begin{align} \label{nx}\phi_-&\leq\phi_n\leq \phi_+\cr
                                      \phi_{n+1}&\leq \phi_n .\end{align}
In this case $\phi_n$ is monotonically decreasing with $n$ and bounded below by $\phi_-$, and so must have a limit for $n\to\infty$.
 The limiting function $\phi=\lim_{n\to\infty}\phi_n$ is clearly bounded by $\phi_-\leq \phi\leq \phi_+$, and satisfies the Lichnerowicz equation,
 since eqn. (\ref{thestep}) converges for large $n$ to $(\Delta_{\t h}-c)\phi=F_c(\phi,x)$.
  
 The inequalities (\ref{nx}) are proved by induction in $n$.   For example, suppose that $\phi_n\leq \phi_+$ for some $n$.   Then
 \begin{align}\label{zp} (\Delta_{\t h}-c)(\phi_{n+1}-\phi_+) &=F_c(\phi_n,x)-\Delta_{\t h}\phi_++c\phi_+ 
 \geq F_c(\phi_n,x)-F(\phi_+,x)+c\phi_+ \cr &=F_c(\phi_n,x)-F_c(\phi_+,x)\geq 0. \end{align}
 The second step holds because $\phi_+$ is a supersolution, and the 
 last step follows from monotonicity of $F_c$.   The maximum principle then implies that $\phi_{n+1}-\phi_+$
 is nonpositive, because if $\phi_{n+1}-\phi_+$ is positive at the point where it achieves its maximum value, then
 $(\Delta_{\t h}-c)(\phi_{n+1}-\phi_+)$ is negative at that point, contradicting eqn. (\ref{zp}).   A similar inductive
 argument proves that $\phi_-\leq \phi_n$ for all $n$.   Finally, to prove inductively that $\phi_{n+1}\leq \phi_n$ for all $n$, one observes that   
 \be\label{hp}(\Delta_{\t h}-c)(\phi_{n+1}-\phi_n)=F_c(\phi_{n},x)-F_c(\phi_{n-1},x). \ee
 By the induction hypothesis $\phi_n\leq \phi_{n-1}$ along with the monotonicity of $F_c$, the right hand side of eqn. (\ref{hp}) is nonnegative.   The same
 argument as before using the maximum principle then implies that $\phi_{n+1}-\phi_n$ is nonpositive.  
 
 This completes the existence proof of the solution of the Lichnerowicz equation on a compact manifold $S$, assuming $R(\t h),\Lambda<0$.   In the AAdS case, one proceeds as follows.   
 Restrict from $S$ to the compact manifold with boundary $S_\epsilon$ defined by $\rho\geq \epsilon$.   The same argument as before, using Neumann boundary
 conditions for the operator $\Delta_{\t h}-c$, produces a solution $\phi_\epsilon$ of the Lichnerowicz equation on $S_\epsilon$ satisfing 
 $\phi_-\leq\phi_\epsilon\leq \phi_+$ and
 also satisfying Neumann boundary conditions on $\partial S_\epsilon$.   In the limit $\epsilon\to 0$, $\phi_\epsilon$ converges to the desired function $\phi$
 that satisfies the Lichnerowicz equation throughout $S$ and approaches 1 at infinity.
 
 To show uniqueness of the solution, first observe that since we know that the solution exists, we can make a Weyl transformation from the initially assumed metric
 $\t h$ to some other metric $h$ that satisfies the Einstein constraint equation.  Saying that $h$ obeys the Einstein constraint equation is equivalent
 to saying that, with background metric $h$, the Lichnerowicz equation is satisfied with $\phi=1$:
 \be\label{pilon} F(1,x)=0. \ee
Now let us ask whether some other function $\phi$ satisfies the Lichnerowicz equation:
 \be\label{zilon}\Delta_h \phi -F(\phi,x)=0. \ee
If so, then
 \begin{align}\label{jilon} 0=&\Delta_h\phi-F(\phi,x)+\phi F(1,x)\cr =
 & \Delta_h\phi +\frac{(d-2)}{4(d-1)}|\t K|^2_{\t h} (\phi^{(2-3d)/(d-2)} -\phi)+\frac{\Lambda(d-2)}{2(d-1)}(\phi^{(d+2)/(d-2)}
 -\phi). \end{align}
This equation implies that $\phi\leq 1$ everywhere, since if the maximum of $\phi$ is at a point $p$ at which $\phi>1$, then each term on the right hand side
is negative at that $p$, which is not possible.  Likewise the equation implies that $\phi\geq 1$ everywhere, since if the minimum of $\phi$ is at a point $p$
at which $\phi<1$, then each term on the right hand side is positive at $p$, again not possible.   So we must have $\phi=1$ and the solution is unique.

This completes our discussion of the Lichnerowicz equation for pure gravity.

\subsection{Symplectic Point Of View}\label{symplectic}

A more conceptual understanding of the Einstein
momentum constraint and its Weyl invariance requires a few steps.\footnote{The following remarks are equally valid in a closed 
universe or open universe and require no assumption about the cosmological constant.
In the AAdS case, of course, one must place appropriate conditions on the behavior of the metric and canonical momentum at infinity, and on the allowed
behavior at infinity of a Weyl transformation or diffeomorphism.}  The canonical momentum in General Relativity is
\be\label{koneq} \Pi^{ij}=\frac{1}{8\pi G} \sqrt{\det h}\left(K^{ij}-\K h^{ij}\right). \ee   The canonical commutation relations between the metric $h$ 
and the canonical momentum $\Pi$ can be summarized by the symplectic form 
\be\label{toneq}\omega =\int_S \largedelta \Pi^{ij}\largedelta h_{ij}, \ee
where $\bigdelta$ is the exterior derivative acting on the infinite-dimensional space $\WW$ of pairs $\Pi, h$ (in finite dimensions, we denote the exterior derivative
as $\d$ rather than $\bigdelta$).  We have
\be\label{koneg}\omega=\largedelta \lambda\ee
with 
\be\label{joneq} \lambda =\int_S \Pi^{ij} \largedelta h_{ij}. \ee   
In classical mechanics, analogous formulas $\lambda=\sum_a p_a\d q^a$, $\omega=\d\lambda$ hold for any classical phase space that is a cotangent bundle
 $T^*Q$, where $Q$ is parametrized by the $q^a$ and the $p_a$ parametrize the fiber directions in the cotangent bundle.   So in the case
of gravity, the full phase space $\WW$, prior to imposing any constraint, is $T^*\M$, where $\M$ is the space of metrics $h$ on $S$.

Two interesting groups  act on this phase space, and we will want to construct reduced phase spaces by imposing these groups as constraints.
    First we observe that the symplectic form $\omega$ and the 1-form $\lambda$ have an obvious Weyl symmetry:
\be\label{hubcap} \largedelta h_{ij}=2\varphi h_{ij},~~~ \largedelta \Pi^{ij}=-2\varphi\Pi^{ij}.\ee    
Let $\W$ be the group of Weyl transformations, and $\Ca=\M/\W$ the space of conformal structures  on $S$.
The Hamiltonian function that generates the Weyl transformation (\ref{hubcap}) by Poisson brackets 
is\footnote{The label ``$g$'' in $\mu_{\varphi,g}$ is for gravity; later we will consider matter contributions
to the Hamiltonian functions.} 
\be\label{kop}\mu_{\varphi,g} =-2\int_S \varphi h_{ij}\Pi^{ij}.\ee   
By setting $\mu_{\varphi,g}=0$ and dividing by $\W$, one can construct a ``reduced phase space,'' called  the symplectic quotient of $\WW$ by $\W$.
Setting  $\mu_{\varphi,g}=0$ means taking $h_{ij}\Pi^{ij}=0$, so that $\Pi$ becomes traceless and  the definition (\ref{koneq}) of $\Pi$  reduces to 
\be\label{wonteq} \Pi^{ij}=\frac{1}{8\pi G} \sqrt{\det h} K^{ij}, \ee
with $K^{ij}$ now constrained by $\K=0$.   After imposing this condition, we  divide by Weyl transformations, acting as in eqn. (\ref{hubcap}).
In terms of $h$ and $K$, the Weyl transformations are
\be\label{ubcap} \largedelta h_{ij}=2\varphi h_{ij},~~~~\largedelta K^{ij} =-2(1+d/2)\varphi K^{ij}.\ee
Thus the reduced phase space, denoted $T^*\M//\W$, is the space of pairs $h,K$, with $K$ traceless, subject to this action of $\W$.
What is described in eqn. (\ref{ubcap}), though written at the Lie algebra level, is 
the same Weyl transformation law for $h$ and $K$ that was introduced previously in eqns. 
 (\ref{olygo}), (\ref{polygo}).   
  In particular,
this derivation gives a better understanding of the possibly mysterious-looking exponent in eqn. (\ref{polygo}).   Since the action of $\W$ on $T^*\M$ comes
from an action on the base space $\M$, the reduced phase space is again a cotangent bundle $T^*\M//\W=T^*(\M/\W)=T^*\Ca$.

Another natural group that acts on these spaces is the diffeomorphism group $\D$ of $S$.  The Lie algebra of $\D$ consists of vector fields on $S$.
The transformation of $h$ generated by a vector field $U$ on $S$ is $\bigdelta h_{ij}=D_i U_j+D_j U_i$.
The Hamiltonian function that generates this transformation is 
   \be\label{muu}\mu_{U,g}= \int_S \Pi^{ij}(D_i U_j+D_j U_i). \ee
   To construct the symplectic quotient $T^*\M/\D$, we set $\mu_{U,g}=0$ and divide by $\D$.   Integrating by parts in eqn. (\ref{muu}), we see that the condition
 that $\mu_{U,g}=0$ for all $U$ is satisfied if and only if 
 \be\label{mino}0= D_i \Pi^{ij}=\frac{1}{8\pi G} \sqrt h \left(D_i K^{ij}-D^j \K\right). \ee
This is the momentum constraint of General Relativity.
 
Diffeomorphisms and Weyl transformations together generate a group that is a semidirect product
$\W\rtimes \D$.     In particular, $\D$ is a group of outer automorphisms of $\W$.  This group structure implies that a Weyl transformation shifts  $\mu_{U,g}$ by a multiple of $\mu_{\phi,g}$; in other words, $\mu_{U,g}$ is Weyl-invariant, once we impose $\mu_{\phi,g}=0$.   This gives the promised conceptual explanation of the
fact that the momentum constraint is Weyl-invariant when restricted to $\K=0$.

To construct the symplectic quotient $T^*\M//\W\rtimes\D$, we have to set $\mu_\varphi=\mu_U=0$ and divide by $\W\rtimes \D$.
In other words $T^*\M//\W\rtimes\D$ parametrizes pairs $K,h$, where $K$ is traceless and obeys the momentum constraint equation, up to equivalence
under diffeomorphisms and Weyl transformations.   Since the action of $\W\rtimes \D$ on $T^*\M$    comes from an action on $\M$,
the symplectic quotient $T^*\M//\W\rtimes\D$ is a cotangent bundle $T^*\P$, where $\P=\M/(\W\rtimes \D)=\Conf/\Diff.$        

What we learned in section  \ref{phase} is that  it is equivalent to impose the Einstein Hamiltonian constraint equation on a pair $K,h$, where $K$ is traceless,
or to divide by Weyl transformations.    So assuming the existence of maximal volume hypersurfaces (so that $K$ can be assumed traceless),
 the phase space of General Relativity in an AAdS spacetime is the cotangent bundle $T^*\P=T^*(\Conf/\Diff)$.

\subsection{Generalization To Include Matter Fields}\label{matf}

It is pleasantly straightforward to generalize what was explained about the Einstein constraint equations in section \ref{phase} to encompass any of the usual
models of gravity coupled to matter fields that satisfy the strong energy condition.   The important examples include scalar fields (possibly forming a nonlinear sigma-model)
with a nonpositive potential energy and $p$-form gauge fields for $p\geq 1$ (possibly generalized to Yang-Mills fields if $p=1$).    Incorporation of such fields in the Lichnerowicz equation has been discussed in 
\cite{I,crusc,Sak,IB}, among other references.

As a first example, we will consider scalar fields.   To simplify the notation,
we consider a single scalar field $\sigma$; the generalization to several scalar fields does not change anything essential.  Assuming that $\sigma$ is canonically normalized, its stress tensor is
\be\label{stressful} T_{ij}=\partial_i\sigma \partial_j\sigma-\frac{1}{2} g_{ij} \partial_k\sigma \partial^k\sigma-\frac{1}{2}g_{ij} V(\sigma). \ee
Here $g$ is the metric on $X$; its restriction to $S$ will be denoted as $h$.
The cosmological constant is included in $V(\sigma)$ as an additive constant.  The strong energy condition is satisfied if and only  $V(\sigma)\leq 0$
 for all $\sigma$.

In studying gravity coupled to a scalar field in AAdS spacetime, we assume that $\sigma$ has a constant value near infinity.  Moreover,
we assume that this constant value is an extremum of $V(\sigma)$, with a negative value of $V$, corresponding to an AdS vacuum.

The phase space of the scalar field $\sigma$ can be parametrized by the restriction of $\sigma$ to an initial value surface $S$ together with a canonical momentum $\pi$.
The symplectic form for this data is 
\be\label{symsigma}\omega_\sigma=\int_S\largedelta \pi \largedelta \sigma =\largedelta\lambda_\sigma, \ee
with
\be\label{lambsig}\lambda_\sigma=\int_S\pi \largedelta \sigma. \ee
To incorporate these variables in the analysis of the Lichnerowicz equation, the first step is to decide how Weyl transformations act on $\pi,\sigma$.
The only general procedure that makes sense is to take $\sigma$ to be Weyl-invariant, since in a general model of scalar fields, $\sigma$ is really the pullback
to spacetime of a function on the target space of a nonlinear sigma-model; in that generality a non-trivial Weyl transformation law for $\sigma$ would not be meaningful.
Once we decide that $\sigma$ is Weyl-invariant, invariance of $\lambda_\sigma$ and $\omega_\sigma$ means that $\pi$ must be Weyl-invariant as well.

In a coordinate system that takes the standard form (\ref{metform}) near $S$, the standard formula for $\pi$ is
\be\label{pkio} \pi =\sqrt h\dot \sigma,\ee
where $\dot\sigma=\partial\sigma/\partial t$.  Since we take Weyl transformations to act on $h$ by $\bigdelta h_{ij}=2\varphi h_{ij}$, eqn. (\ref{pkio}), together with the
Weyl-invariance of $\pi$, implies that $\dot\sigma$ must transform as $\bigdelta \dot\sigma =-d\dot\sigma$,   In sum,
\be\label{kio}\largedelta\sigma=0, ~~~\largedelta\dot\sigma=-d\dot\sigma. \ee

The Einstein momentum constraint equation with the field $\sigma$ included and with $K$ assumed to be traceless is 
\be\label{zymsigma}   D_i K^{ij}=8\pi G T^{0j}=-8\pi G \dot\sigma \partial_k\sigma h^{kj}. \ee
This equation is Weyl-invariant, with Weyl transformations taken to act by eqns. (\ref{hubcap}) and (\ref{pkio}).  
This Weyl invariance may come as a slight surprise, but it has the same explanation as in section \ref{symplectic} in terms of a symplectic quotient. 
To see this, let $\Sigma$ be the infinite-dimensional space that parametrizes the values of the field $\sigma$ on $S$.  Then the phase space of $\sigma$ is
$T^*\Sigma$, where the fiber directions are parametrized by $\pi$. So   $\sigma$, $h$, and their canonical momenta jointly parametrize $T^*(\M\times\Sigma)$.
Let us consider the symplectic quotient of this phase space by the group $\W\rtimes\D$.    First  we need to compute the contributions of $\sigma$ to
the Hamiltonian functions $\mu_\varphi$ and $\mu_U$.    The contribution to $\mu_{\varphi}$ vanishes
because $\sigma$ and $\pi$ are Weyl-invariant.     So setting $\mu_{\varphi}=0$ will mean setting $\K=0$,
just as in the absence of  $\sigma$.   On the other hand, $\sigma$ does contribute to $\mu_U$.   The contribution is\footnote{Here $\iota_V$ is the operation of
contracting the first index of a differential form with a vector field $V$.} 
\be\label{thec} \mu_{U,\sigma}=-\iota_{V_U}\lambda_\sigma
=-\int_S \pi U^k\partial_k\sigma =-\int_S \sqrt h \dot\sigma U^k\partial_k\sigma. \ee
So the condition for vanishing of the total Hamiltonian function $\mu_U=\mu_{U,g}+\mu_{U,\sigma}$ is
\be\label{new} -D_i K^{ij}=8\pi G \dot\sigma \partial_k\sigma h^{jk},\ee
which is the Einstein momentum constraint for this coupled system.   The same group theoretic considerations as before imply that the momentum constraint
is Weyl-invariant, once we set $\K=0$.

To construct the symplectic quotient $T^*(\M\times \Sigma)//\W\rtimes\D$, we set $\K=0$, impose the Einstein momentum constraint, and divide
by $\W\rtimes \D$.   Since the action of $\W\rtimes\D$ on $T^*(\M\times\Sigma)$ is induced in the usual way from an action on $\M\times \Sigma$,
the result is a cotangent bundle $T^*\P_\Sigma$, where $\P_\Sigma =(\M\times\Sigma)/(\W\rtimes\D)$  parametrizes pairs $(h,\sigma)$
up to diffeomorphism and Weyl transformation. 

To construct the phase space of this system, we would follow all of the same steps except that instead of dividing by $\W$, we would impose the Einstein Hamiltonian
constraint equation
\be\label{phogo} R=K_{ij} K^{ij}+16\pi G T_{00}. \ee
However, essentially the same arguments as summarized in section \ref{phase} shows that it is equivalent to impose the Hamiltonian constraint or to divide
by the group of Weyl transformations, since each orbit of $\W$ contains a unique point at which the Hamiltonian constraint equation is satisfied.   

To show this, we consider the orbit of $\W$ that contains a set of fields $\t h, \t K, \t\sigma,\dot{\t\sigma}$.    This can be Weyl-transformed to
\begin{align}\label{ohmy} h & = \phi^{4/(d-2)}\t h \cr
                                              K^{ij}&= \phi^{-2(d+2)/(d-2)}\t K^{ij} \cr
                                                \sigma & = \t \sigma \cr
                                                \dot\sigma& = \phi^{-2d/(d-2)}\dot{\t\sigma},\end{align}
                                                where $\phi$ is an arbitrary positive function.   The energy density is
\begin{align}\label{zolbo}T_{00}=&\frac{1}{2}\dot\sigma^2+\frac{1}{2}\partial_i\sigma \partial_j \sigma h^{ij}+V(\sigma)   \cr
   = &     \frac{1}{2}\phi^{-4d/(d-2)}\dot{\t\sigma}^2+\frac{1}{2}\phi^{-4/(d-2)} \partial_i\t\sigma \partial_j \t\sigma \t h^{ij}+V(\t\sigma)  .\end{align}                                                                 
Repeating the derivation that led to eqn. (\ref{poneq}), we get the new form of the Lichnerowicz equation.
This is actually an equation of the same general form as before, but with new coefficients:
\be\label{newc}   \Delta_{\t h} \phi -\frac{(d-2)}{4(d-1)} \alpha \phi+\frac{(d-2)}{4(d-1)}\beta \phi^{(2-3d)/(d-2)} +\frac{(d-2)}{4(d-1)}\gamma\phi^{(d+2)/(d-2)}=0,\ee
with
\begin{align}\label{longlist} \alpha& = R(\t h)-8\pi G \partial_i\t \sigma \partial_j\t\sigma \t h^{ij} \cr
                                           \beta&=|\t K|_{\t h}^2+16\pi G \t{\dot\sigma}^2\cr
                                            \gamma& =16\pi G V(\t\sigma). \end{align}
                                            
In section \ref{phase}, to prove the existence of a solution of the Lichnerowicz equation, we needed\footnote{The definition of a subsolution and a supersolution
actually requires these statements
to hold uniformly, so that for example, instead of just saying that $\alpha<0$, we need to have a positive constant $\epsilon$ such that $\alpha<-\epsilon$ everywhere.   
On a noncompact
manifold, in general such a uniform inequality might be a stronger condition than $\alpha<0$, but with AAdS boundary conditions (and $\sigma$ assumed to be
constant at infinity) the two statements are equivalent.}
$\alpha<0$, $\gamma<0$.   The proof of uniqueness of the solution
required $\beta>0$, $\gamma<0$.         Incorporating the scalar field $\sigma$ does not affect the conditions $\alpha<0$ and $\beta>0$, and it does not affect the condition
$\gamma<0$ in a model that satisfies the strict strong energy condition $V(\sigma)<0$ for all $\sigma$.   Thus, in such a model, the Lichnerowicz equation has a unique
positive 
solution that approaches 1 at infinity.   This assertion is actually Theorem 3.3 in \cite{Sak}.

Under these conditions, solving the Hamiltonian constraint equation has the same effect as dividing by the group of Weyl transformations, and therefore the part of the phase
space that  parametrizes spacetimes that can be described by a solution of the Lichnerowicz equation with $\K=0$ is a cotangent bundle $T^*\P_\Sigma$.    
We argued in section     \ref{exi}
that given reasonable (but optimistic) assumptions about singularities in General Relativity, every solution has a maximal Cauchy hypersurface and hence can
be described by a $\K=0$ solution of the Lichernowicz equation.   Thus under this assumption, the phase space is $T^*\P_\Sigma$.      

The slightly more general case of  a model that satisfies  $V(\sigma)\leq 0$ everywhere 
but not necessarily   $V(\sigma)<0$ is analyzed in  \cite{Sak}, Theorem 7.1.

\subsection{$p$-Form Gauge Fields}\label{pform}

In this section, we will generalize from a scalar field to a $p$-form gauge field\footnote{For $p=0$, a $p$-form gauge field is the same 
as a scalar field, already discussed in section \ref{matf}.  For $p=1$, a $p$-form gauge field is an abelian gauge field.   Abelian gauge theories can be generalized to
 nonabelian gauge theories, but this generalization does not affect our considerations here.}
 $A$ with gauge transformation $A\to A+\d\lambda$, for a $(p-1)$-form $\lambda$, and with  $(p+1)$-form field strength $F=\d A$.   Here $0\leq p\leq D-2$.     
As usual, if $S$ is a Cauchy hypersurface, we can pick local coordinates such that $S$ is defined by $t=0$ and near $S$ the metric takes the form
\be\label{nears}\d s^2=-\d t^2+h_{ij}(\vec x,t)\d x^i \d x^j.\ee
Along $S$,
we  decompose $F=B+\d t \wedge E$, where the ``magnetic'' field $B$ is a $(p+1)$-form along $S$, and the ``electric'' field $E$ is a $p$-form along $S$.

We have two goals in studying a $p$-form gauge field: (1) to show that the standard theories of a $p$-form gauge field satisfy the 
strong energy condition; (2) to incorporate such a field in the analysis of the Einstein constraint equations.

With the usual normalization, the stress tensor of a minimally coupled $p$-form gauge field is
\be\label{zolgo} T_{\mu\nu}=\frac{1}{p!} F_{\mu\alpha_1\cdots \alpha_p}F_\nu\,^{\alpha_1\cdots \alpha_p} -\frac{1}{2(p+1)!} g_{\mu\nu} F_{\alpha_0\alpha_1\cdots\alpha_p}
F^{\alpha_0\alpha_1\cdots \alpha_p}. \ee
From this, we can compute
\be\label{tzz} T_{00}=\frac{1}{2p!} E_{i_1 \cdots i_p}E^{i_1\cdots i_p}+\frac{1}{2(p+1)!} B_{i_0 i_1\cdots i_p}B^{i_0i_1\cdots i_p}. \ee
Likewise
\begin{align}\label{nzz}T^\alpha_\alpha &=\frac{1}{p!}\left(1-\frac{D}{2(p+1)}\right) F_{\alpha_0\cdots \alpha_p}F^{\alpha_0\cdots\alpha_p} 
\cr & = \frac{1}{p!}\left(1-\frac{D}{2(p+1)}\right) \left(-(p+1)  E_{i_1\cdots i_p}E^{i_1\cdots i_p}+ B_{i_0 i_1\cdots i_p} B^{i_0 i_1\cdots i_p} \right). \end{align}
Remembering the definition $\t T_{\mu\nu}=T_{\mu\nu}-\frac{1}{D-2}g_{\mu\nu} T^\alpha_\alpha$, we find
\be\label{welfind} \h T_{00}= \frac{1}{(D-2)p!}\left( (D-p-2) E_{i_1 \cdots i_p}E^{i_1\cdots i_p} + p  B_{i_0 i_1\cdots i_p} B^{i_0 i_1\cdots i_p} \right). \ee
This is manifestly non-negative in the whole range $0\leq p\leq D-2$, showing that these theories satisfy the strong energy condition.

Several commonly studied nonminimal couplings of a $p$-form gauge field do not affect this analysis.   Chern-Simons couplings do not contribute to the stress
tensor, so they do not affect the strong energy condition.  A $p$-form gauge field can couple to a scalar field $\phi$ in such a way that the action is, for example,
$\int e^\phi F\wedge \star F$ rather than the minimal $\int F\wedge \star F$.  This merely multiplies the stress tensor by $e^\phi$, without effect on the above
analysis.   One can also have Higgsing of a $p$-form gauge field by a $(p-1)$-form gauge field (as a result of which the $p$-form gauge field becomes massive).  
This again does not disturb the analysis.

Bearing in mind these comments, we see that eleven-dimensional supergravity, and also the Type I, Type IIA, and Type IIB supergravities in ten dimensions,
all satisfy the strong energy condition.    However,  massive Type IIA supergravity \cite{Romans} does not satisfy the strong energy condition, since it has a scalar
field with a positive potential.\footnote{Nonetheless,  massive Type IIA supergravity does satisfy the Maldacena-Nu\~{n}ez
no go theorem \cite{MN} for de Sitter compactifications.}  

Now we will discuss the incorporation of these fields in the Einstein constraint equations.  First, we have to decide how Weyl transformations should act in this theory.

The gauge invariance $A\to A+\d \lambda$ would not intertwine with a nontrivial Weyl transformation law for $A$ in any reasonable way, so $A$ must be Weyl-invariant.
Hence $B=\d A$ is Weyl-invariant.   Another way to reach the same conclusion is to observe that the Bianchi identity satisfied by $B$, and the quantized Dirac fluxes that
it can carry, would not be consistent
with any nontrivial Weyl transformation law for $B$. So $B$ must be Weyl-invariant.  

Since $A$ is Weyl-invariant, its canonical momentum $\Pi$ must also be Weyl-invariant.    With the usual normalization, $\Pi^{i_1i_2 \cdots i_p}=\sqrt h E^{i_1 i_2
\cdots i_p}$.     So $E^{i_1 i_2\cdots i_p}$ must transform as $1/\sqrt h$, and equivalently $E_{i_1 i_2\cdots i_p}$ 
must transform as $h^{p-d/2}$.   This tells us the $p$-form analog of eqn. (\ref{ohmy}):
\begin{align}\label{nohmy} h & = \phi^{4/(d-2)}\t h \cr
                                              K^{ij}&= \phi^{-2(d+2)/(d-2)}\t K^{ij} \cr
                                                B_{i_0\cdots i_p} & = \t B_{i_0\cdots i_p} \cr
                                                E_{i_1\cdots i_p} & = \phi^{(4p-2d)/(d-2)}   \t E_{i_1\cdots i_p}.\end{align}
                                                
Because $A$ and $\Pi$ are Weyl-invariant, the $p$-form gauge field, just like the scalar field studied in section \ref{matf}, does not contribute to the Hamiltonian
generator $\mu_\varphi$ of Weyl transformations.  So we remain with $\mu_\varphi=\K$, and imposing Weyl-invariance as a constraint means setting $\K=0$ and then
dividing by Weyl transformations,
exactly as before.   Once we set $\K=0$, the momentum constraint equation, which now has a contribution proportional to the momentum density $T^{0i}$ of the
$p$-form gauge field, again becomes Weyl-invariant.       The group theoretic explanation for this fact is exactly as before.     

It remains to examine the contribution of the $p$-form gauge field to the Lichnerowicz equation.   From eqns. (\ref{tzz}) and (\ref{nohmy}), we find for this field
\be\label{fondue} T_{00}= \frac{\phi^{-4(d-p)/(d-2)}}{2p!}  |\t E|^2_{\t h} +\frac{\phi^{-4(p+1)/(d-2)} }{2(p+1)!} |\t B|^2_{\t h}. \ee
As a check, note that this is consistent with the usual duality under $p\leftrightarrow d-p-1$ with exchange of $E$ and $B$.    
Following the familiar steps, the Lichnerowicz equation comes out to be 
\begin{align} \label{newlich} &\Delta_{\t h} \phi -\frac{(d-2)}{4(d-1)} R(\t h) \phi+\frac{(d-2)}{4(d-1)}|\t K|^2_{\t h} \phi^{\frac{(2-3d)}{(d-2)}} 
+\frac{\Lambda(d-2)}{2(d-1)}\phi^{\frac{(d+2)}{(d-2)}}     \cr &+ \frac{4\pi G(d-2)}{(d-1)}\left( \frac{1}{2 p!} \phi^{\frac{(-3d+4p+2)}{(d-2)}} |\t E|^2_{\t h}
          +\frac{1}{2(p+1)!} \phi^{\frac{(d-4p-2)}{(d-2)}} |\t B|^2_{\t h}    \right)=0. \end{align}
 The method of subsolutions and supersolutions  applies exactly as before to show the existence of a solution of this equation with $\phi\to 1$ at infinity.      
 All that we need to know is that the $|\t E|^2_{\t h}$ and $|\t B|^2_{\t h}$ terms that have been added are positive, so they do not change the sign  of
 the left hand side of the equation if $\phi$ is  small and 
 constant, and they are subdominant for large $\phi$, so they also do not change the sign if $\phi$ is large and constant. 
 
 The functions of $\phi$ that multiply $|\t E|^2_{\t h}$ and $|\t B|^2_{\t h}$ in eqn. (\ref{newlich}) are of the form $\phi^\alpha$ where $\alpha\leq 1$ for all
 $p$ in the range $0\leq p\leq d-1$.   This bound on the exponent ensures that the additional terms in the equation do not affect the proof of uniqueness of the
 solution of the Lichnerowicz equation, which proceeds  as in the discussion of eqn. (\ref{jilon}).

\subsection{$\AdS$ Compactifications}\label{compactification}

The analysis of the Lichnerowicz equation works so nicely for a spacetime that is asymptotic to $\AdS_D$ for some $D$ that it perhaps comes as a surprise
that compactification to $\AdS_D$ is different.  In other words, if we consider a spacetime that is asymptotic to $\AdS_D\times W$ for some compact manifold
$W$ of positive dimension, we do not get such a simple picture.

As a typical example, we will consider solutions of ten-dimensional Type IIB supergravity that are asymptotic at infinity to $\AdS_5\times S^5$.   Type IIB supergravity
has a four-form gauge field $A$ whose five-form field strength $F=\d A$ is self-dual.  In the standard $\AdS_5\times S^5$ solution of Type IIB supergravity,
$F$ is everywhere nonzero.   Type IIB supergravity has bosonic fields other than the metric and $A$, but including them would not qualitatively change the picture, so
for brevity we omit them.  

We can find the Lichnerowicz equation appropriate to a Type IIB spacetime $X$ that is asymptotic to $\AdS_5\times S^5$ by setting $d=9$, $p=4$ in eqn.
(\ref{newlich}).  Self-duality of the five-form $F$ means that the $|\t E|^2_{\t h}$ and $|\t B|^2_{\t h}$ terms in the equation are equal, and we actually
should keep only one of them.    We also have to set $\Lambda=0$, since Type IIB supergravity in ten dimensions has vanishing cosmological constant.  So
the Lichnerowicz equation becomes
\begin{align} \label{tenlich} &\Delta_{\t h} \phi -\frac{(d-2)}{4(d-1)} R(\t h) \phi+\frac{(d-2)}{4(d-1)}|\t K|^2_{\t h} \phi^{-27/7} 
    + \frac{4\pi G(d-2)}{2(d-1)(p+1)!} \phi^{-9/7} |\t B|^2_{\t h}    =0. \end{align}
    
To proceed, we need to take into account one more key fact.   In studying  the Lichnerowicz equation on a manifold asymptotic to $\AdS_D$, we always required
$R(\t h)$ to be negative at infinity.   However, in a spacetime $X$ that is asymptotic to $\AdS_5\times S^5$, we instead want $R(\t h)$ to be positive near infinity. 
One way to see this is to observe that the usual $\AdS_5\times S^5$ spacetime actually has zero scalar curvature, since the negative scalar curvature of $\AdS_5$ is 
equal and opposite to the positive scalar curvature of $S^5$.   However, if we restrict to a Cauchy hypersurface $S$ such as $\AdS_4\times S^5$, then
as the scalar curvature of $\AdS_4$ is less negative than that of $\AdS_5$, we see that $S$ actually has positive scalar curvature.  

More directly, we can look at the Hamiltonian constraint equation, which for an extremal  hypersurface $S$ reads
\be\label{hamc} R(h)=K^{ij}K_{ij} +16\pi G \h T_{tt}. \ee
Since Type IIB supergravity satisfies the strong energy condition, as observed in section \ref{pform}, we will have $\h T_{tt}\geq 0$ everywhere in a solution
of this theory, and therefore any extremal hypersurface will always have $R(h)\geq 0$ everywhere.    Note that this argument is not in any way special
to $\AdS_5\times S^5$; it applies to any AAdS compactification of eleven-dimensional supergravity or of Type IIA, Type IIB, or Type I supergravity in ten
dimensions, since these models all satisfy the strong energy condition, as found in section \ref{pform}.  Because of this, what we are explaining here applies
to a very wide range of Anti de Sitter compactifications, though we consider $\AdS_5\times S^5$ for illustration.

For the standard extremal Cauchy hypersurface $S=\AdS_4\times S^5$ in the standard $\AdS_5\times S^5$ spacetime, $B$ is everywhere nonzero
and the scalar curvature $R(h)$ is everywhere positive.   Therefore, for any pair $\t h, \t B$  sufficiently close to this standard example, $\t B$ is everywhere nonzero
and  $R(\t h)$ is everywhere positive.    Under this restriction, the analysis of the Lichnerowicz equation actually
proceeds rather as before, with minor differences.   The left hand side of the equation
is positive for $\phi$ a small positive  constant and negative for large constant $\phi$, though
the negativity for large $\phi$ now comes from the fact that $R(\t h)$ is
assumed positive rather than from having $\Lambda<0$.   Given this property of the equation, the method of subsolutions and supersolutions applies to prove
the existence of a solution $\phi$ of the equation with $\phi\to 1$ at infinity.   In addition, the various powers of $\phi $ appearing in the equation are such that
the solution of the Lichnerowicz equation is unique, by the same argument as in eqn. (\ref{jilon}).   

What happens in the case of a solution that is not close to the standard example? Of the two assumptions that we made in getting to this point, the assumption that $B$ is everywhere nonzero is relatively harmless, since $B$ has
$9!/4!5!=126$ components, and generically is everywhere nonzero in nine dimensions.   However, the assumption that $R(\t h)$ is everywhere
positive is highly problematic.  When we studied spacetimes asymptotic to $\AdS_D$, we used the fact that it is always possible, by a Weyl transformation,
to find a starting point with $R(\t h)<0$.   But in studying compactifications to Anti de Sitter space, we would want to make a Weyl transformation to set $R(\t h)>0$.
This is not always possible.

In fact, there are strong topological obstructions to the existence on a manifold $M$ of a metric of positive scalar curvature.   The simplest obstruction is as follows.
Type IIB supergravity has fermions, so we can assume that $S$ is a spin manifold.  Therefore it has a Dirac operator $\DD= \i\slashed{D}$.  A spin manifold
of dimension $8k+1$ has a topological  invariant known as the ``mod 2 index'' \cite{AS}, the number of zero modes of the Dirac operator mod 2.    
A nine-dimensional spin manifold $M$
with a nonzero mod 2 index cannot admit a metric of positive scalar curvature.   In fact, the square of the Dirac operator, namely  $\DD^2=-D_\mu D^\mu+R/4$,
is strictly positive if $R>0$, so $\DD$  has no zero-modes on a manifold of positive scalar curvature \cite{Lichn}. Therefore, on a manifold $M$ on which the ordinary index  of the Dirac operator (if $M$ has dimension $4k$)  or
the mod 2 index (if $M$ has dimension  $8k+1$ or $8k+2$) is nonzero, there is no metric of positive scalar curvature.  To be more precise, such results are usually
stated and proved on a compact manifold $M$, and we are interested in a noncompact nine-manifold $S$.  However, we want a complete metric on $S$ such
that the scalar curvature approaches a positive constant at infinity (namely the scalar curvature of $\AdS_4\times S^5$).   This, together with the formula 
$\DD^2=-D_\mu D^\mu+R/4$, implies   that the Dirac operator
on $S$ has a discrete spectrum near 0.  Given this, positivity of $R$ implies that the index and the mod 2 index must vanish, just as on a compact manifold.

The formula $\DD^2=-D_\mu D^\mu +R/4$ also shows that $\DD^2$ is strictly positive if $R\geq 0$ everywhere and $R$ is not identically 0.   In our application,
we are interested in metrics for which $R$ is strictly positive near infinity and in particular not identically zero.   A nonzero mod 2 index implies that there is no such
metric with $R\geq 0$ everywhere.

A simple example of a nine-manifold with a nonzero mod 2 index is provided by a certain exotic nine-sphere.   In general, roughly speaking, half of all
nine-dimensional spin manifolds have a nonzero mod 2 index.   Actually,  the mod 2 index is only the simplest example of an obstruction to positive
scalar curvature.   A more systematic study \cite{Gro} shows that, roughly speaking, most manifolds with a large fundamental group do not admit a metric
of positive scalar curvature.   On the other hand, for a simply-connected nine-dimensional spin manifold, the mod 2 index is the only obstruction to 
having a metric of positive scalar curvature \cite{Stolz}.    Again, such results
are most often stated for compact manifolds but apply equally to, for example, nine-manifolds that are asymptotic to $\AdS_4\times S^5$ with the stipulation
that the scalar curvature should approach a positive constant at infinity.

Now consider a spacetime $X$ that is asymptotic to $\AdS_5\times S^5$ and has a Cauchy hypersurface $S$ that, topologically,
does not admit a metric of positive scalar curvature.   The Hamiltonian constraint equation (\ref{hamc}) implies immediately that if it is possible to choose
$S$ to have $\K=0$, then the scalar curvature of $S$ is nonnegative.   Thus, if $X$ is such that a Cauchy hypersurface  $S\subset X$ has a nonzero mod 2 index,
then it is not possible to choose such an $S$ to satisfy $\K=0$.   

On the other hand, if $S$ is any nine-manifold asymptotic to $\AdS_4\times S^5$, there is no problem to find initial data on $S$ that satisfy the Einstein constraint
equations if we relax the assumption $\K=0$.   When $\K\not=0$, there is an additional term $-\K^2$ on the right hand side of the Hamiltonian constraint equation,
and there is no reason to expect that $R\geq 0$ everywhere.

Therefore, there are perfectly good spacetimes $X$  asymptotic to $\AdS_5\times S^5$ and completely generic but not possessing any extremal Cauchy hypersurface.
What are we to make of this?   Based on the discussion in section \ref{exi}, though the arguments are not truly bullet-proof, we suspect that in such a spacetime, some sort of unfamiliar, non-BKL singularity forms
generically.

We will speculate in a moment on how this might be interpreted, but first let us note that in the context of an asymptotically flat spacetime, the obstruction we are
discussing  to
the existence of an extremal slice was discovered long ago \cite{Brill}.  The original context
for this work was that it had been conjectured that in an asymptotically flat spacetime $X$, for any value of the time measured at infinity, there would be an extremal
Cauchy slice $S$ in the interior of $X$; it was shown by considering topological obstructions to positive scalar curvature
that this is not the case.  On the other hand, it was found that by allowing $\K\not=0$, 
one can find initial data leading to an asymptotically flat spacetime $X$ with any assumed topology of $S$ \cite{Wittold,Witt}.   

We have simply pointed out precisely the same topological obstruction
 in the context of compactifications to Anti de Sitter space.  However, the implications are somewhat different.
In an asymptotically flat spacetime $X$, the domain of dependence of a Cauchy hypersurface $S$ is all of $X$, and  is never compact.   However, in, for example,
a spacetime $X$ that is asymptotic to $\AdS_5\times S^5$, the domain of dependence $\Omega$ of a Cauchy hypersurface  $S$ is compact in the
absence of singularities. This was explained in section \ref{exi}.  So the  potential connection between the topological obstruction to $R\geq 0$ and 
 singularities is special to AdS compactifications.

What are we to say about these hypothetical non-BKL singularities?   We can only make some speculative remarks.  As an example, consider the mod 2 index
as an obstruction to positive scalar curvature.   It is an invariant in spin bordism, which means that from the point of view of classical physics, if one assumes
that the relevant spacetime histories are smooth manifolds  (possibly not admitting
 a metric everywhere of Lorentz signature, as discussed for example in \cite{SorLouk}), the mod 2 index is a conserved quantity.   There is no corresponding
$\Z_2$ gauge field, so this is a candidate as a global conserved $\Z_2$ charge.   On the other hand, one does not expect global conservation laws in quantum
gravity (for the most precise known argument for  this assertion, see \cite{HO}), and in particular we do not expect cobordism invariants to be truly conserved
\cite{VCob}.  So we expect that there is some sort of process in Type IIB superstring theory in which the mod 2 index changes.   Perhaps the non-BKL singularity
that is suggested by the arguments we have sketched is a signal of such topology change.  One can imagine a singularity that arises when a  topological
defect of some sort that  supports the mod 2 index collapses to a point and disappears.  There is certainly no known singularity in General Relativity associated to such a time-dependent process, so if this type of topology change is associated to a singular classical history, this is a classical history with a singularity of an unknown and exotic type.

More generally, a nine-dimensional spin manifold $S$ has many possible topologies, but one expects that most topological distinctions between different
initial value surfaces are not well-defined in the full Type IIB superstring theory -- even though only a few special cases of topology-changing processes
are  well-understood.   It is possible that the non-BKL singularities that the analysis here suggests play a role in filling in the gaps and providing missing
topology-changing processes.   

We conclude with the following remarks.   In asymptotically $\AdS_D$ spacetimes, in models that satisfy the strong
energy condition, we learned from the study of the Lichnerowicz equation that -- with an optimistic but not obviously wrong assumption about the nature of singularities  --
the gravitational phase space is a cotangent bundle, with potential implications for quantization.   On the other hand, in the context of a spacetime $X$ asymptotic
to $\AdS_D\times W$, for some compact $W$, this is not the case:  if a Cauchy hypersurface $S\subset X$ is such that positive scalar curvature is
topologically obstructed, then there is a perfectly good phase space of classical solutions of Type IIB supergravity with this $S$, but there is no reason
for it to be a cotangent bundle.

In fact, even if $S$ is such that there is no obstruction to positive scalar curvature, we cannot prove that the phase space is a cotangent bundle, no matter
what assumption we make about possible singularities.   The reason  is that even if $S$ does admit metrics with $R\geq 0$ and appropriate
asymptotic behavior, this does not mean that every metric on $S$ with appropriate asymptotic behavior is Weyl-equivalent to one with $R\geq 0$ everywhere.
For example, if $S$ is topologically $\AdS_4\times S^5$, we only know {\it a priori} that any metric sufficiently close to the standard one has $R\geq 0$.   (In fact, it is possible
to prove that there are metrics on $S$ with the desired behavior at infinity that are not conformal to any metric with $R\geq 0$ everywhere.) 

So in  spacetimes asynptotic to $\AdS_D\times W$, as opposed to $\AdS_D$, the phase space is not going to be a cotangent bundle for each topological choice
of initial value surface, even with optimistic assumptions.  However, we know little about the generic singularities in these spacetimes.   At the cost of going rather
far out on a limb, we can speculate that perhaps different classical
 phase spaces associated to spacetimes with different topologies, after taking singularities and topology-changing
processes and massive stringy modes into account, do fit together to make a cotangent bundle.   

The singularities associated to topology change might conceivably be mild enough that a hypersurface $S$ can be sensibly continued from one side of the
singularity to the other.   In that case, it might  be that ultimately  the maximal volume hypersurface does always exist, but in general on a spacetime with a
different topology than what we started with.

\vskip1cm
 \noindent {\it {Acknowledgements}}  Research supported in part by NSF Grant PHY-2207584.   I thank F. Bonsante,  P. T. Chru\'{s}ciel, G. Galloway, M. Henneaux, G. Horowitz, D. Jafferis, J. Maldacena,
   D. Marolf, R. Mazzeo, and J.-M. Schlenker for discussions.
 \bibliographystyle{unsrt}

\end{document}